\documentclass[11pt]{article}
\usepackage{epsfig}
\begin{document}
\input amssym.def 
\input amssym
\hfuzz=5.0pt
%
%
%
%
\def\vec#1{\mathchoice{\mbox{\boldmath$\displaystyle\bf#1$}}
{\mbox{\boldmath$\textstyle\bf#1$}}
{\mbox{\boldmath$\scriptstyle\bf#1$}}
{\mbox{\boldmath$\scriptscriptstyle\bf#1$}}}
\def\mbf#1{{\mathchoice {\hbox{$\rm\textstyle #1$}}
{\hbox{$\rm\textstyle #1$}} {\hbox{$\rm\scriptstyle #1$}}
{\hbox{$\rm\scriptscriptstyle #1$}}}}
\def\operatorname#1{{\mathchoice{\rm #1}{\rm #1}{\rm #1}{\rm #1}}}
\chardef\ii="10
\def\widehat{\mathaccent"0362 }
\def\widetilde{\mathaccent"0365 }
\def\vphi{\varphi}
\def\vrho{\varrho}
\def\vtheta{\vartheta}
\def\ih{{\i\over\hbar}}
\def\hi{\frac{\hbar}{\i}}
\def\CD{{\cal D}}
\def\CE{{\cal E}}
\def\CH{{\cal H}}
\def\CL{{\cal L}}
\def\CP{{\cal P}}
\def\CV{{\cal V}}
\def\CQ{{\cal Q}}
\def\fo{{\frak o}}
\def\fs{{\frak s}}
\def\half{{1\over2}}
\def\bhalf{\hbox{$\half$}}
\def\viert{{1\over4}}
\def\bviert{\hbox{$\viert$}}
\def\bsechszehn{\hbox{$\frac{1}{16}$}}
\def\hhbox#1#2{\hbox{$\frac{#1}{#2}$}}
\def\dfrac#1#2{\frac{\displaystyle #1}{\displaystyle #2}}
\def\intT{\ih\int_0^\infty\d\,T\,e^{\i ET/\hbar}}
\def\pathint#1{\int\limits_{#1(t')=#1'}^{#1(t'')=#1''}\CD #1(t)}
\def\hbarm{{\dfrac{\hbar^2}{2m}}}
\def\hbarmq{{\dfrac{\hbar^2}{2mq}}}
\def\mzwei{\dfrac{m}{2}}
\def\overh{\dfrac1\hbar}
\def\ihbar{\dfrac\i\hbar}
\def\intt{\int_{t'}^{t''}}
\def\tn{\tilde n}
\def\pmb#1{\setbox0=\hbox{#1}
    \kern-.025em\copy0\kern-\wd0
    \kern.05em\copy0\kern-\wd0
    \kern-.025em\raise.0433em\box0}
\def\pathintG#1#2{\int\limits_{#1(t')=#1'}^{#1(t'')=#1''}\CD_{#2}#1(t)}
\def\limN{\lim_{N\to\infty}}
\def\Norm{\bigg({m\over2\pi\i\epsilon\hbar}\bigg)}
\def\hbaram{{\hbar^2\over8m}}
\def\tlist#1{$\underline{\hbox{#1}}$}
\def\sdrei{{S^{(3)}}}
\def\SdreiC{{S_{3C}}}
\def\ldrei{{\Lambda^{(3)}}}
\def\hhbox#1{\hbox{$#1$}}
\def\Partial#1{\frac{\partial}{\partial #1}}
\def\Partialsq#1{\frac{\partial^2}{\partial{#1^2}}}
\def\bbbr{{\rm I\!R}}                                
\def\bbbn{{\rm I\!N}}                                
\def\bbbz{{\mathchoice {\hbox{$\sf\textstyle Z\kern-0.4em Z$}}
{\hbox{$\sf\textstyle Z\kern-0.4em Z$}}
{\hbox{$\sf\scriptstyle Z\kern-0.3em Z$}}
{\hbox{$\sf\scriptscriptstyle Z\kern-0.2em Z$}}}}    
\def\bbbc{{\mathchoice {\setbox0=\hbox{\rm C}\hbox{\hbox
to0pt{\kern0.4\wd0\vrule height0.9\ht0\hss}\box0}}
{\setbox0=\hbox{$\textstyle\hbox{\rm C}$}\hbox{\hbox
to0pt{\kern0.4\wd0\vrule height0.9\ht0\hss}\box0}}
{\setbox0=\hbox{$\scriptstyle\hbox{\rm C}$}\hbox{\hbox
to0pt{\kern0.4\wd0\vrule height0.9\ht0\hss}\box0}}
{\setbox0=\hbox{$\scriptscriptstyle\hbox{\rm C}$}\hbox{\hbox
to0pt{\kern0.4\wd0\vrule height0.9\ht0\hss}\box0}}}}
\def\Energyldrei{\exp\bigg[-{\i\hbar T\over2m}(p^2+1)\bigg]}
\def\Me{\operatorname{Me}} 
\def\U{\operatorname{U}} 
\def\Ai{\operatorname{Ai}} 
\def\Cl{\operatorname{Cl}} 
\def\ee{\operatorname{e}} 
\def\oo{\operatorname{o}} 
\def\OO{\operatorname{O}} 
\def\so{\operatorname{so}} 
\def\SO{\operatorname{SO}} 
\def\SU{\operatorname{SU}} 
\def\SS{\operatorname{S}} 
\def\dt{\d t}
\def\d{\,\operatorname{d}}
\def\e{\,\operatorname{e}}
\def\i{\operatorname{i}}
\def\max{\,\operatorname{max}}
\def\sign{\,\operatorname{sign}}
\def\cn{\,\operatorname{cn}}
\def\dn{\,\operatorname{dn}}
\def\sn{\,\operatorname{sn}}
\def\ps{\,\operatorname{ps}}
\def\diag{\operatorname{diag}}
\def\Emetric{(k^2\cn^2\alpha+{k'}^2\cn^2\beta)}
\def\emetric{k^2\cn^2\alpha+{k'}^2\cn^2\beta}
\def\vphi{\varphi}
\def\operatorname#1{{\mathchoice{\rm #1}{\rm #1}{\rm #1}{\rm #1}}}
\def\bbbone{{\mathchoice {\rm 1\mskip-4mu l} {\rm 1\mskip-4mu l}
{\rm 1\mskip-4.5mu l} {\rm 1\mskip-5mu l}}}
\def\pathint#1{\int\limits_{#1(t')=#1'}^{#1(t'')=#1''}\CD #1(t)}
\def\pathints#1{\int\limits_{#1(0)=#1'}^{#1(s'')=#1''}\CD #1(s)}

\begin{titlepage}
\centerline{\normalsize DESY 07--133 \hfill ISSN 0418 - 9833}
\centerline{\hfill August 2007}
\vskip.3in
\message{TITLE:}
\begin{center}
{\Large Path Integral Representations on the Complex Sphere}
\end{center}
\message{Path Integral Representations on the Complex Sphere}
\vskip.5in
\begin{center}
{\Large Christian Grosche}
\vskip.2in
{\normalsize\em II.\,Institut f\"ur Theoretische Physik}
\vskip.05in
{\normalsize\em Universit\"at Hamburg, Luruper Chaussee 149}
\vskip.05in
{\normalsize\em 22761 Hamburg, Germany}
\end{center}
\normalsize
\vfill
\begin{center}
{\bf Abstract}
\end{center}
In this paper we discuss the path integral representations for the coordinate 
systems on the complex sphere $\SdreiC$. The Schr\"odinger equation, 
respectively the path integral, separates in exactly 21 orthogonal coordinate 
systems. We enumerate these coordinate systems 
and we are able to present the path integral representations explicitly in the
majority of the cases. In each solution the expansion into the wave-functions 
is stated. Also, the kernel and the corresponding Green function can be stated 
in closed form in terms of the invariant distance on the sphere, respectively
on the hyperboloid. 
\end{titlepage}
 
\tableofcontents

\setcounter{page}{1}
\setcounter{equation}{0}
\section{Introduction}
\message{Introduction}
It is a well-known fact that the number of coordinate systems which separate 
the Schr\"odinger equation on the three-dimensional sphere is six: cylindrical,
spherical, conical, prolate spheroidal, oblate spheroidal, and elliptic
\cite{GROaa,GROad}. On the two-dimensional sphere there are just two,
spherical and elliptical. 
On the other hand, the corresponding number of coordinate systems 
on the two- and three-dimensional hyperboloid are nine, respectively 34
\cite{OLE}. Furthermore, on the two- and three-dimensional Euclidean and
pseudo-Euclidean spaces the number of systems is 10 and 54, and 4 and 11,
respectively. 
All the spaces listed above can have an Euclidean metric signature, or a
Minkowskian metric signature, and they are all real.
The situation changes, if we start to consider the corresponding complex
spaces. On two-dimensional and three-dimensional complex Euclidean space there
are 6 and 18, respectively, and on the two- and three-dimensional complex
sphere there are 5 and 21 coordinate systems which separate 
the Schr\"odinger equation \cite{BKWb}. In comparison to the real two- and
three-dimensional sphere there is a richer structure. This is not surprising
because the two-dimensional flat space and the hyperboloids are contained as
subgroup cases in the complex sphere.

\begin{table}[t!]
\caption{Coordinate Systems on $\SdreiC$}
\vspace{-0.5cm}
\label{SdreiCcosys}
\begin{eqnarray}\begin{array}{l}\vbox{\small\offinterlineskip
\halign{&\vrule#&$\strut\ \hfil\hbox{#}\hfill\ $\cr
\noalign{\hrule}
height2pt&\omit&&\omit&&\omit&&\omit&&\omit&&\omit&\cr
&No.\ of  &&Complexifcation  &&Two-Dimensional Subsystem  
                             &&Three-Dimensional Systems &\cr
&System   &&of Real Space    &&  
                             && &\cr
height2pt&\omit&&\omit&&\omit&&\omit&&\omit&&\omit&\cr
\noalign{\hrule}\noalign{\hrule}
height2pt&\omit&&\omit&&\omit&&\omit&&\omit&&\omit&\cr
&No 1  &&$S^{(3)},\Lambda^{(3)}$, O(2,2)	 
&& $S(2,\bbbr)$ Polar          &&Cylindrical         &\cr
&No 2  &&$\Lambda^{(3)}$, O(2,2)
&& $E(2,\bbbr)$ Cartesian      &&$\hbox{\underline{Horicyclic}}$  &\cr
&No 3  &&$S^{(3)},\Lambda^{(3)}$, O(2,2)
&& $S(2,\bbbr)$ Polar          &&Spherical           &\cr
&No 4  &&$\Lambda^{(3)}$, O(2,2)
&& $S(2,\bbbc)$ Horospherical  &&$\hbox{\underline{Horospherical}}$&\cr
&No 5  &&$\Lambda^{(3)}$, O(2,2)
&& $E(2,\bbbr)$ Polar          &&$\hbox{\underline{Horicyclic-polar}}$    &\cr
&No 6  &&$S^{(3)},\Lambda^{(3)}$, O(2,2)
&& $S(2,\bbbr)$ Conical        &&Sphero-elliptic     &\cr
&No 7  &&$\Lambda^{(3)}$, O(2,2)
&& $S(2,\bbbc)$ Degenerate elliptic I               
       &&Spherical-degenerate elliptic I                    &\cr
&No 8  &&$\Lambda^{(3)}$, O(2,2)
&& $S(2,\bbbc)$ Degenerate elliptic II               
       &&Spherical-degenerate elliptic II                   &\cr
&No 9  &&$\Lambda^{(3)}$, O(2,2)
&& $E(2,\bbbr)$ Elliptic       &&$\hbox{\underline{Horicyclic-elliptic}}$ &\cr
&No 10 &&O(2,2)
&& $E(2,\bbbc)$ Hyperbolic     &&$\hbox{\underline{Horicyclic-hyperbolic}}$&\cr
&No 11 &&$\Lambda^{(3)}$, O(2,2)
&& $E(2,\bbbr)$ Parabolic      &&$\hbox{\underline{Horicyclic-parabolic I}}$
                                                                          &\cr
&No 12 &&O(2,2)
&& $E(2,\bbbc)$ Semi-parabolic &&$\hbox{\underline{Horicyclic-parabolic II}}$ 
                                                                          &\cr
height2pt&\omit&&\omit&&\omit&&\omit&&\omit&&\omit&\cr
\noalign{\hrule}\noalign{\hrule}
height2pt&\omit&&\omit&&\omit&&\omit&&\omit&&\omit&\cr
&No 13 &&$S^{(3)},\Lambda^{(3)}$, O(2,2)
&& -   &&$\hbox{\underline{Elliptic-Cylindrical}}$&\cr
&No 14 &&$\Lambda^{(3)}$, O(2,2)
&& -                           &&Elliptic-Parabolic  &\cr
&No 15 &&$\Lambda^{(3)}$, O(2,2)
&& -                           &&Elliptic-Hyperbolic &\cr
&No 16 &&O(2,2)
&& -   &&$\hbox{\underline{Parabolic}}$           &\cr
&No 17 &&$S^{(3)},\Lambda^{(3)}$, O(2,2)
&& -                           &&Ellipsoidal      &\cr
&No 18 &&$\Lambda^{(3)}$, O(2,2)
&& -                           &&{\it System 18}&\cr
&No 19 &&O(2,2)
&& -                           &&{\it System 19}&\cr
&No 20 &&$\Lambda^{(3)}$, O(2,2)
&& -                           &&{\it System 20}&\cr
&No 21 &&O(2,2)
&& -                           &&{\it System 21} &\cr
height2pt&\omit&&\omit&&\omit&&\omit&&\omit&&\omit&\cr
\noalign{\hrule}}}\end{array}\nonumber\end{eqnarray}
\vspace{-0.5cm}
\end{table}

In Table \ref{SdreiCcosys} I have listed some properties of the coordinate
systems on the three-dimensional complex sphere. 
The coordinate systems, which contain two-dimensional flat systems, i.e. the
Euclidean plane (real and complex) are, (2), (5), (9)--(11) \cite{MILLf}.
The coordinates systems which have the two-dimensional sphere (real and
complex) as a subsystem are (3), (4), (6)--(8) \cite{KAL,KAMIi,KMP1,MILLf}.
The coordinate systems which exist also on the real three-dimensional sphere
are (1), (3), (6), (13) and (17) \cite{KMW,OLE}.

According to \cite{KAMIj}, the complexification of the two elliptic cylindrical
coordinate systems (i.e. spheroidal systems) on the $S(3,\bbbr)$-sphere and on
the three-dimensional hyperboloid give just one coordinate system on
$S(3,\bbbc)$, i.e.~(13).  In particular in \cite{MPST-A,KMP1} coordinate systems on
the two-dimensional complex sphere and corresponding superintegrable
potentials, and in \cite{KMP1} coordinate systems on the two-dimensional
(complex) plane and corresponding superintegrable potentials were discussed,
including the corresponding interbases expansions.
The goal of \cite{KMP1} was to extend the notion of superintegrable
potentials of real spaces to the corresponding complexified spaces.
The findings were such that there are in addition to the four coordinate systems on
the real two-dimensional Euclidean plane three more coordinate systems and
also three more superintegrable potentials.
Similarly, in addition to the two coordinate systems on the real
two-dimensional sphere there are three more coordinate systems on the complex
sphere and also four more superintegrable potentials.
This is not surprising because the complex plane contains not only the
Euclidean plane but also the pseudo-Euclidean plane (10 coordinate systems
\cite{GROad,KAL,KALc}) and the complex sphere contains not only 
the real sphere but also the two-dimensional hyperboloid (9 coordinate systems
\cite{GROad,KAL,KAMIb,OLE}). 

On the other hand, it is also possible to complexify the 34 coordinate systems
on the three-dimensional hyperboloid \cite{KAMIc,OLE}. 
This gives all 21 systems except (19, (12), (16), (19) and (21). 
These remaining system can be found be complexifying the coordinate
systems on the real hyperboloid $\SO(2,2)$ \cite{KAMIj}.

In Table \ref{SdreiCcosys} I have indicated which coordinate systems emerges
from which real space by complexification; these spaces are indicated by 
$S^{(3)}$ (real sphere), $\Lambda^{(3)}$ (real hyperboloid),
and the systems emerging from O(2,2) (real hyperboloid), respectively.

In this paper I apply the path integral method \cite{FH,KLEo,GRSh,SCHUH} to
the complex sphere $\SdreiC$. We are able to find in the majority 
of the coordinate representations  a path integral representation.
Many known solutions from other path integral problems can be applied in a
straightforward way to find the corresponding complex-sphere representation. 

In Table \ref{SdreiCcosys} those coordinate systems 
where a new path integral representation can be stated are 
$\underline{\hbox{underlined}}$.
The most important ``new'' solution consist in the path integral formulation
of the complex periodic Liouville potential. It has a real spectrum with 
eigenvalues $\propto J(J+2), J\in\bbbn$, in accordance with the spectrum of the
Hamiltonian of the three-dimensional sphere. The fact that a complex periodic
potential has a real spectrum is at first sight surprising, but has actually
attracted in recent years a lot of attention, e.g. the workshop series 
\cite{Znojil}. This property of ``pseudo-hermitian'' Hamiltonians has also 
become known as ``${\cal P}{\cal T}$-symmetry'' of Hamiltonians.

In Table \ref{SdreiCcosys} those ellipsoidal systems where no path integral
evaluation is possible are {\it emphasized}. There is hardly a solution 
known, because the {\it only} ellipsoidal systems where a solution of the 
free Schr\"odinger is know is the one on the real sphere \cite{AKPSZ}. 
There is nothing known about the solutions of the Schr\"odinger equations 
on the hyperboloids in paraboloidal or the remaining ``ellipsoidal''
coordinates, let alone a path integral representation.

The paper has a review-like character and is organized as follows:
In the next Section we present the relevant path integral
representations for those coordinate systems which have a subgroup structure,
i.e. where a lower dimensional case is contained. In many cases we can rely 
on already solved path integral problem to find the specific one in question.
Section III contains the remaining cases which are not of the subspace type. 
For some evaluations I also will make use of the technique of space-time
transformation, where I refer for more details to the literature 
\cite{DK,GRSb,GRSh,KLEo}. Therefore our principal emphasis is to bring
together both already known and new results in order to complete matters.

Obviously, Sections II and III take on the form of an enumeration.
We will rely heavily on already known solutions in the sequel, and these
solution will not be re-derived again.

We start with the definition of the coordinates, calculate the
relevant metric terms, the momentum operator and the quantum Hamiltonian 
in terms of he momentum operators.
In our formulation of the path integral we always use a lattice definition
which we have called ``product form'' and consists mainly of the fact that 
all metric terms in the energy term in the Lagrangian in the its lattice form
are given by geometric means \cite{GRSh}. In our first path integral
representation we will state this lattice formulation explicitly.

Let us note that the complexification requires also the following
consideration:  The eigenvalues of the
Hamiltonian on the three-dimensional complex sphere are denoted by
$\propto-\sigma(\sigma+2)$. On the real sphere this yields with $\sigma =J$
the eigenvalues $\propto J(J+2)$ whereas on the real three-dimensional 
hyperboloid one has $\sigma=1+\i p$ and therefore the eigenvalues are 
$\propto (p^2+1)$. We must therefore look carefully which manifold we 
consider if we specify the coordinates including their ranges.
Usually, an analytic continuation may be required, which is not performed, however.

We will denote in the following the quantum number of the Eigenvalues
of the Hamiltonian by $J$, irrespective whether there is a discrete or a
continuous spectrum. On the real sphere the
spectrum is always discrete and on the hyperboloid (two-sheeted) continuous.
The kernel in its wave-functions expansion and spectrum on the complex sphere
will be displayed in the most cases in a discrete formulation. The
corresponding restriction to 
the sphere or hyperboloid then will decide whether one can keep the discrete
formulation as it is or one must analytically continue to the continuous
spectrum. However, on the single-sheeted hyperboloid and on the
O(2,2)-hyperboloid one has in fact both: a discrete and a continuous part.
The latter is not discussed in the sequel and is postponed to a future study.
In the enumeration of the coordinate system we keep the convention of the
corresponding systems on the sphere and hyperboloid from previous
publications \cite{GROad,OLE}. 
In some cases we note explicitly the correspondences to the
sphere and hyperboloid cases to illustrate the examples.

The third section contains a summary and discussion of our results.


\setcounter{equation}{0}
\section{The Path Integral Representations: Part I}
\message{The Path Integral Representations: Part I}

\subsection{System 1: Cylindrical }
\message{System 1: Cylindrical }
The cylindrical coordinate system on the complex sphere $\SdreiC$ has
the form:
\begin{equation}
\left.\begin{array}{ll}
z_1=\sin\vtheta\cos\vphi_1 &z_2=\sin\vtheta\sin\vphi_1\\
z_3=\cos\vtheta\cos\vphi_2 &z_4=\cos\vtheta\sin\vphi_2
\end{array}\qquad\right\}\qquad
(\vtheta\in(0,\hbox{$\frac{\pi}{2}$}),\vphi_1,\vphi_2\in[0,2\pi))\enspace.
\end{equation}
The set of commuting operators characterizing the cylindrical
coordinate system are
\begin{equation}
\CL_1=I_{23}^2, \qquad\CL_2=I_{14}^2\enspace.
\end{equation}
The metric terms have the form
\begin{equation}
\left.\begin{array}{l}
\d s^2=\d\vtheta^2+\sin^2\vtheta\d\vphi_1^2+\cos^2\vtheta\d\vphi_2^2\enspace,
\\[3mm]
\sqrt{g}=\sin\vtheta\cos\vtheta\enspace,\\[3mm]
\Gamma_\vtheta=\cot\vtheta-\tan\vtheta,\quad
\Gamma_{\vphi_1}=0,\quad \Gamma_{\vphi_2}=0\enspace.
\end{array}\qquad\right\}
\end{equation}
Therefore we have for the momentum operators:
\begin{equation}
p_\vtheta=\hi\bigg(\Partial{\vtheta}+\half\cot\vtheta-\half\tan\vtheta\bigg),
\qquad p_{\vphi_1}=\hi\Partial{\vphi_1},\qquad
p_{\vphi_2}=\hi\Partial{\vphi_2},
\end{equation}
and the Hamiltonian is given by
\begin{eqnarray}  
H&=&-\hbarm\Bigg[\Partialsq{\vtheta}
                  +(\cot\vtheta-\tan\vtheta)\Partial{\vtheta}
     +\frac{1}{\sin^2\vtheta}\Partialsq{\vphi_1}
     +\frac{1}{\cos^2\vtheta}\Partialsq{\vphi_2}\Bigg]
\nonumber\\   &=&
    \frac{1}{2m}\Bigg(p_{\vtheta}^2+\frac{1}{\sin^2\vtheta}\,p_{\vphi_1}^2
+\frac{1}{\cos^2\vtheta}\,p_{\vphi_2}^2\Bigg)
-\hbaram\bigg(4+{1\over\cos^2\vtheta}
                    +{1\over\sin^2\vtheta}\bigg)\enspace. 
\label{Hamiltonian}
\end{eqnarray}  
In the second line in (\ref{Hamiltonian}) a quantum potential $\propto\hbar^2$
appears which is due to the ordering of the position and momentum operators in $H$.
The path integral solution on the complex sphere $\SdreiC$ is
identical to the the usual sphere $S_{3\bbbr}$ and is well-known from the
literature and I just state the result. 
\newline
In the canonical formulation we have \cite{GROab,GROad,GRSb,GRSh}
(compare also \cite{BJb,CAMP,SCHUHa})
\begin{eqnarray}       & &\!\!\!\!\!\!\!
K^{(\SdreiC)}(\vtheta'',\vtheta',\vphi_1'',\vphi_1',\vphi_2'',\vphi_2';T)
         \nonumber\\   & &\!\!\!\!\!\!\!
=\lim_{N\to\infty}\Norm^{3N/2}\prod_{j=1}^{N-1}
 \int\d\theta_j\sin\vtheta_j\cos\vtheta_j\d\vphi_{1,j}d\vphi_{2,j}
         \nonumber\\   & &\!\!\!\!\!\!\!\qquad\times
\exp\Bigg\{\ih\sum_{j=1}^N\Bigg[\frac{m}{2}\Big(
\Delta^2\vtheta_j\!+\!\widehat{\cos^2\vtheta_j}
\Delta^2\vphi_{1,j}\!+\!\widehat{\sin^2\vtheta_j}\Delta^2\vphi_{2,j}\Big)\!
+\!\hbaram\bigg(4\!+\!{1\over\cos^2\vtheta_j}
                    \!+\!{1\over\sin^2\vtheta_j}\bigg)\Bigg]dt\Bigg\}
         \nonumber\\   & &\!\!\!\!\!\!\!
 =\pathint\vtheta\sin\vtheta\cos\vtheta\pathint{\vphi_1}\pathint{\vphi_2}
         \nonumber\\   & &\!\!\!\!\!\!\!\qquad\times
  \exp\Bigg\{\ih\intt\Bigg[{m\over2}\Big(
\dot\vtheta^2+\cos^2\vtheta\dot\vphi_1^2+\sin^2\vtheta\dot\vphi_2^2\Big)
  +\hbaram\bigg(4+{1\over\cos^2\vtheta}
                    +{1\over\sin^2\vtheta}\bigg)\Bigg]dt\Bigg\}
         \nonumber\\   & &\!\!\!\!\!\!\!
  =\sum_{J=0}^\infty\sum_{k_1,k_2\in\bbbz}
   {\e^{\i[k_1(\vphi_1''-\vphi_1')+k_2(\vphi_2''-\vphi_2')]}\over4\pi^2}
   \frac{2(J+1)\left(\frac{J-|m_1|-|m_2|}{2}\right)!
                  \left(\frac{J+|m_1|+|m_2|}{2}\right)!}
        {\left(\frac{J-|m_1|+|m_2|}{2}\right)!
         \left(\frac{J+|m_1|-|m_2|}{2}\right)!}
        \nonumber\\   & &\!\!\!\!\!\!\!\qquad\times\vphantom{\bigg]}
  (\sin\vtheta'\sin\vtheta'')^{|m_1|}
  (\cos\vtheta'\cos\vtheta'')^{|m_2|}
        \nonumber\\   & &\!\!\!\!\!\!\!\qquad\times\vphantom{\bigg]}
  P_{(J-|m_1|-|m_2|)/2}^{(|m_1|,|m_2|)}(\cos2\vtheta')
  P_{(J-|m_1|-|m_2|)/2}^{(|m_1|,|m_2|)}(\cos2\vtheta'')
  \,\e^{-\i\hbar TJ(J+2)/2m}
\end{eqnarray}
($u_j=u(t_j)$, $\Delta u_j=u_j-u_{j-1}$, $u_j=u(t'+t_j)$, 
for $u=\vtheta,\vphi_1,\vphi_2$, $\epsilon=T/N$, $\widehat{\sin^2\vtheta_j}
=\sin^2\vtheta_j\sin^2\vtheta_{j-1}$, and for $\cos^2\vtheta$ similarly). 
Here we have stated the explicit lattice definition of the path integral. 
Thus we have for the wave-functions on $\SdreiC$ and the energy-spectrum
\begin{eqnarray}
&&\Psi_{J,m_1,m_2}(\vtheta,\vphi_1,\vphi_2)=
\sqrt{\frac{2(J+1)\left(\frac{J-|m_1|-|m_2|}{2}\right)!
                  \left(\frac{J+|m_1|+|m_2|}{2}\right)!}
        {\left(\frac{J-|m_1|+|m_2|}{2}\right)!
         \left(\frac{J+|m_1|-|m_2|}{2}\right)!}}
\nonumber\\   & &\qquad\qquad\qquad\qquad\times
{\e^{\i(k_1\vphi_1+k_2\vphi_2)}\over2\pi}
  (\sin\vtheta)^{|m_1|}(\cos\vtheta)^{|m_2|}
  P_{(J-|m_1|-|m_2|)/2}^{(|m_1|,|m_2|)}(\cos2\vtheta)\enspace,\qquad
\\
&&E_J=\frac{\hbar^2}{2m}J(J+2)\enspace.
\label{Energy-J}
\end{eqnarray}
The kernel $K^{(\sdrei)}(T)$ can be cast into the form of a $\Theta$-function
\cite{GRA}
\begin{equation}
K^{(\sdrei)}(\psi_\sdrei,T)=
  {\e^{\i\hbar T/2m}\over4\pi^2}{\d\over\d\cos\psi_\sdrei}
   \Theta_3\Bigg({\psi_\sdrei\over2}\bigg|-{\hbar T\over2\pi m}\Bigg)\enspace.
\label{KS3}  
\end{equation}
The corresponding Green function (resolvent kernel) has the form
\begin{equation}
G^{(\sdrei)}(\psi_\sdrei,E)={m\over2\pi\hbar^2}
  {\sin[(\pi-\psi_\sdrei)(\gamma+1/2)]\over
   \sin[\pi(\gamma+1/2)]\sin\psi_\sdrei}\enspace,
\label{GS3}  
\end{equation}
where the quantity $\cos\psi_\sdrei$ (invariant distance) in spherical 
coordinates is given by
\begin{equation}
  \cos\psi_\sdrei(\vec z'',\vec z')=\cos\vtheta_1'\cos\vtheta_1''
  +\sin\vtheta_1'\sin\vtheta_1''\Big(\cos\vtheta_2'\cos\vtheta_2''
  +\sin\vtheta_2'\sin\vtheta_2''\cos(\vphi''-\vphi')\Big)\,,
\label{distance-sphere}
\end{equation}
and $\gamma=-1/2+\sqrt{2mE/\hbar^2+1}$.
The expressions (\ref{KS3}) and (\ref{GS3}) are
independent of the particular coordinate representation.

Note that the corresponding representation of the Green function of the 
three-dimensional hyperboloid has the form 
($K_{\Lambda^{(3)}}(T)$ cannot be stated in closed
form \cite{GROad})
\begin{equation}
G_{\Lambda^{(3)}}(d_\ldrei(\mbf u'',\mbf u'),E)
={-m\over\pi^2\hbar^2\sinh d_\ldrei(\mbf u'',\mbf u')}
  \CQ^{1/2}_{-1/2-\i {\sqrt{2mE/\hbar^2-1}}}
   \Big(\cosh d_\ldrei(\mbf u'',\mbf u')\Big)\enspace,
\label{Green-hyperboloid}
\end{equation}
$\cosh d$ is the invariant distance on $\ldrei$ given by
\begin{equation}
  \cosh d_\ldrei(\mbf u'',\mbf u')=\cosh\tau'\cosh\tau''
  -\sinh\tau'\sinh\tau''\Big(\cos\vtheta'\cos\vtheta''
  +\sin\vtheta'\sin\vtheta''\cos(\vphi''-\vphi')\Big)\,,
\label{distance-hyperboloid}
\end{equation}
with $\tau,\vtheta,\vphi$ spherical coordinates on $\Lambda^{(3)}$.
$\CQ$ denotes a Legendre function of the second kind.
The cylindrical system exists on the three-dimensional sphere (System I.) 
and on the three-dimensional hyperboloid (System I., 
in the following we use the notations and enumerations of \cite{GROad,OLE}).
Note the difference in the Legendre-function (first or second kind) depending
whether we consider the real sphere or the real hyperboloid, respectively.

\subsection{System 2: Horicyclic }
\message{System 2: Horicyclic }
The horicyclic coordinate system on the complex sphere $\SdreiC$ has
the form:
\begin{equation}
\left.\begin{array}{ll}
 z_1=\half\Big[\e^{-\i x}+(1+y^2+z^2)\e^{\i x}\Big]\qquad
&z_2=\i y\e^{\i x}\\
 z_3=\i z\e^{\i x}
&z_4=\frac{\i}{2}\Big[\e^{-\i x}+(1+y^2+z^2)\e^{\i x}\Big]
\end{array}\qquad\right\}
\end{equation}
$(x,y,z\in\bbbr$). The set of commuting operators characterizing the
horicyclic coordinate system are
\begin{equation}
\CL_1=(I_{42}+\i I_{21})^2, \qquad\CL_2=(I_{34}+\i I_{13})^2\enspace.
\end{equation}
The metric terms have the form
\begin{equation}
\left.\begin{array}{l}
\d s^2=\d x^2+\e^{2\i x}(\d y^2+\d z^2)\enspace,
\\[3mm]
\sqrt{g}=\e^{2\i x}\enspace,\\[3mm]
\Gamma_x=2\i,\quad
\Gamma_{y}=0,\quad \Gamma_{z}=0\enspace.
\end{array}\qquad\right\}
\end{equation}
This coordinate system corresponds to a subgroup algebra of $\fs\fo(4,\bbbc)$,
namely ${\cal E}(2,\bbbc)$. According to Ref.\cite{MILLf} there exist six
such systems (which make up the six coordinate systems on 
${\cal  E}(2,\bbbc)$), and together with the horicyclic systems they are
systems (5), (9)--(12). 
The momentum operators are:
\begin{equation}
p_x=\hi\bigg(\Partial{x}+\i\bigg),
\qquad p_{y}=\hi\Partial{y},\qquad
p_{z}=\hi\Partial{z}.
\end{equation}
The Hamiltonian is given by
\begin{eqnarray}  
H&=&-\hbarm\Bigg[\Partialsq{x}+2\i\Partial{x}
     +\e^{-2\i x}\bigg(\Partialsq{y}+\Partialsq{z}\bigg)\Bigg]
\nonumber\\   &=&
    \frac{1}{2m}\Bigg[p_{x}^2+\e^{-2\i x}\Big(p_{y}^2+p_{z}^2\Big)\Bigg]
    -\hbarm\enspace.
\end{eqnarray}  
For the path integral we find
\begin{eqnarray}
&&\!\!\!\!\!\!\!\!
K^{(\SdreiC)}(x'',x',y'',y',z'',z':T)
\nonumber\\  &&\!\!\!\!\!\!\!\!
=\pathint{x}\pathint{y}\pathint{z}\e^{2\i x}
\exp\left\{\ih\int_0^T\Bigg[\frac{m}{2}\dot x^2+\e^{2\i x}(\dot y^2+\dot z^2)
+\frac{\hbar^2}{2m}\Bigg]\dt\right\}
\nonumber\\  &&\!\!\!\!\!\!\!\!
=\e^{-\i(x''-x')}\int_{\bbbr}\d k_y\int_{\bbbr}\d k_z
\frac{\e^{\i k_y(y''-y')+\i k_z(z''-z')}}{(2\pi)^2}
\nonumber\\  &&\!\!\!\!\!\!\!\!\qquad\times
\pathint{x}\exp\left\{\ih\int_0^T
\left[\frac{m}{2}\dot x^2-\frac{\hbar^2}{2m}(k_y^2+k_z^2)\e^{-2\i x}
\right]\dt\right\}.
\end{eqnarray}
The path integration in the variables $y$ and $z$ are just plane waves. The 
remaining path integral in the variable $x$
we can solve this path integral by an analytic continuation of the
Liouville path integral solution \cite{GRSh}. In \cite{CJT} it was shown that
the proper continuation are Hankel functions
$2^{-1/2}H_{n+1/2}^{(1)}(k\,\e^{-\i x})$ \cite{GRA}, hence we obtain:
\begin{eqnarray}
&&\!\!\!\!\!\!\!\!\!\!\!\!K^{(\SdreiC)}(x'',x',y'',y',z'',z':T)
\nonumber\\  &&\!\!\!\!\!\!\!\!\!\!\!\!
=\e^{-\i(x''-x')}\int_{\bbbr}\d k_y\int_{\bbbr}\d k_z
\frac{\e^{\i k_y(y''-y')+\i k_z(k_z(z''-z')}}{(2\pi)^2}
\nonumber\\  &&\!\!\!\!\!\!\!\!\!\!\!\!\qquad\times\!
\sum_{J\in\bbbn_0}\! \bhalf
H^{(1)}_{J+1/2}\Big(\sqrt{k_y^2+k_z^2}\,\e^{-\i x''}\Big)
H^{(1)}_{J+1/2}\Big(\sqrt{k_y^2+k_z^2}\,\e^{-\i x'}\Big)
\exp\Bigg[-\ih\frac{\hbar^2J(J+2)}{2m}T\Bigg]\,.\qquad
\end{eqnarray}
The normalization follows from the consideration
$(1/\sqrt{2})H^{(1)}_{n}(k\e^{\i x})\to\e^{\i kx}/\sqrt{2\pi}$
($x\to\infty$).
Note the relation $H^{(1)}_\nu(\i z)=(2/\i\pi)\e^{\i\pi\nu/2}K_\nu(z)$.
The wave-functions on $\SdreiC$  and the energy-spectrum are
\begin{eqnarray}
\Psi_{Jk_yk_z}(x,y,z)&=&\frac{\e^{\i(k_yy+k_zz)}}{2\pi}\cdot
\e^{-\i x}\,\frac{1}{\sqrt{2}}
H^{(1)}_{J+1/2}\Big(\sqrt{k_y^2+k_z^2}\,\e^{-\i x}\Big)\enspace,
\\
E_J&=&\frac{\hbar^2}{2m}J(J+2)\enspace.
\end{eqnarray}
The horicyclic system exists only on the three-dimensional hyperboloid 
(System II.).

\subsection{System 3: Spherical }
\message{System 3: Spherical }
The spherical coordinate system is the best-known coordinate system
and has been discussed extensively in the literature, 
including its generalization to $D$ dimensions. We have:
\begin{equation}
\left.\begin{array}{ll}
 z_1=\sin\chi\cos\vtheta
&z_2=\sin\chi\sin\vtheta\cos\vphi\\
 z_3=\sin\chi\sin\vtheta\sin\vphi
&z_4=\cos\chi
\end{array}\qquad\right\}\qquad
(\chi\in[0,\pi),\vtheta\in[0,\pi),\vphi\in[0,2\pi))\enspace.
\end{equation}
The set of commuting operators is given by
\begin{equation}
\CL_1=I_{12}^2+I_{13}^2+I_{23}^2,\qquad \CL_2=I_{23}^2\enspace.
\end{equation}
The metric terms are given by
\begin{equation}
\left.\begin{array}{l}
\d
s^2=\d\chi^2+\sin^2\chi(\d\vtheta^2+\sin^2\vtheta\d\vphi^2)
\\[3mm]
\sqrt{g}=\sin\vtheta\sin^2\chi\enspace,\\[3mm]
\Gamma_\vtheta=\cot\vtheta,\quad \Gamma_\chi=2\cot\chi,\quad 
\Gamma_\vphi=0\enspace.
\end{array}\qquad\right\}
\end{equation}
The spherical system corresponds to the $\fs\fo(3,\bbbc)$ subalgebra
of $\fs\fo(4,\bbbc)$ such that the subsystem-coordinates are coordinates
of the two-dimensional complex sphere $S_{2C}$:
$z_1^2+z_2^2+z_3^2=1$. According to Refs.\cite{KAL,KAMIi,KMP1} there are
exactly five such systems, and together with the spherical system these are the
systems (4), (6), (7), and (8).
The momentum operators are:
\begin{equation}
p_\chi=\hi\bigg(\Partial{\chi}+\cot\chi\bigg),\quad
p_\vtheta=\hi\bigg(\Partial{\vtheta}+\half\cot\vtheta\bigg),\quad 
p_{\vphi}=\hi\Partial{\vphi}\enspace,
\end{equation}
and the Hamiltonian is given by
\begin{eqnarray}  
H&=&-\hbarm\Bigg[\Partialsq{\chi}+2\cot\chi\Partial{\chi}
+\frac{1}{\sin^2\chi}\bigg(\Partialsq{\vtheta}+\cot\vtheta\Partial{\vtheta}
+\frac{1}{\sin^2\vtheta}\Partialsq{\vphi}\bigg)\Bigg]
\nonumber\\   &=&
    \frac{1}{2m}\Bigg[p_{\chi}^2
    +\frac{1}{\sin^2\chi}\bigg(p_{\vtheta}^2
    +\frac{1}{\sin^2\vtheta}p_{\vphi}^2\Bigg)\Bigg]
    -\hbaram\Bigg(4+{1\over\sin^2\chi}\bigg(1+
            {1\over\sin^2\vtheta}\bigg)\Bigg)\enspace.
\end{eqnarray}  
I do not go into the details of the path integral solution of the
spherical system, because this has been done extensively in the 
literature, e.g.
\cite{BJb,CAMP,GRSb,GRSh,GROab,GROad,SCHUHa}. We state
\begin{eqnarray}       & &\!\!\!\!\!\!\!    
K^{(\SdreiC)}(\chi'',\chi',\vtheta'',\vtheta',\vphi'',\vphi';T)
         \nonumber\\   & &\!\!\!\!\!\!\!
  =\pathint\chi\sin^2\chi\pathint\vtheta\sin\vtheta\pathint\vphi
         \nonumber\\   & &\!\!\!\!\!\!\!\qquad \times
  \exp\Bigg\{\ih\intt\Bigg[{m\over2}\Big(\dot\chi^2
   +\sin^2\chi(\dot\vtheta^2+\sin^2\vtheta\dot\vphi^2)\Big)
  +\hbaram\Bigg(4+{1\over\sin^2\chi}\bigg(1+
            {1\over\sin^2\vtheta}\bigg)\Bigg)\Bigg]dt\Bigg\}
         \nonumber\\   & &\!\!\!\!\!\!\!
  ={1\over2\pi^2}\sum_{J=0}^\infty(J+1)C_J^1(\cos\psi_\sdrei)
  \exp\bigg[-{\i\hbar T\over2m}J(J+2)\bigg]
\label{numEC}     \\   & &\!\!\!\!\!\!\!
  =\sum_{J=0}^\infty\sum_{m_1,m_2}
  \Psi_{J,m_1,m_2}(\chi'',\vtheta'',\vphi'')
  \Psi_{J,m_1,m_2}^*(\chi',\vtheta',\vphi')
  \exp\bigg[-{\i\hbar T\over2m}J(J+2)\bigg]\enspace,
\label{numEc}     \\   & &\!\!\!\!\!\!\!
  \Psi_{J,m_1,m_2}(\chi,\vtheta,\vphi)=N^{-1/2}\e^{\i m_1\vphi}
   (\sin\chi)^{m_1}C_{J-m_1}^{m_1+2}(\cos\chi)
   (\sin\vtheta)^{m_2}C_{m_1-m_2}^{m_2+3/2}(\cos\vtheta)\enspace,
\vphantom{\bigg]} \\   & &\!\!\!\!\!\!\!
  N={2\pi^32^{-(1+2m_1+2m_2)}\over(J+1)(m_1+3/2)(J-m_1)!(m_1-m_2)!)}
 {\Gamma(J+m_1+2)\Gamma(m_1+m_2+1)\over\Gamma^2(m_1+1)\Gamma^2(m_2+3/2)}
\enspace,
\end{eqnarray}
and the energy-spectrum (\ref{Energy-J}).
The spherical system exists on the three-dimensional sphere (System III.) 
and on the three-dimensional hyperboloid (System X.).

\subsection{System 4: Horospherical}
\message{System 4: Horospherical}
This coordinate system is defined as
\begin{equation}
\left.\begin{array}{ll}
\displaystyle
 z_1=\half\Big[\e^{-\i x}+(1-y^2)\e^{\i x}\Big]\sin\chi\qquad
&z_2=y\e^{\i x}\sin\chi\\
\displaystyle
 z_3=-\frac{\i}{2}\Big[\e^{-\i x}-(1+y^2)\e^{\i x}\Big]\sin\chi
&z_4=\cos\chi
\end{array}\qquad\right\}
\end{equation}
($\chi\in(0,\pi),x,y\in\bbbr$).
The set of commuting operators is given by
\begin{equation}
\CL_1=I_{12}^2+I_{13}^2+I_{23}^2,\qquad \CL_2=(I_{32}+\i I_{21})^2\enspace,
\end{equation}
and the metric terms have the form
\begin{equation}
\left.\begin{array}{ll}
\d s^2=\d\chi^2+\sin^2\chi(\d x^2+\e^{2\i x}\d y^2)\enspace,
\\[3mm]
\sqrt{g}=\e^{\i x}\sin^2\chi\enspace,\\[3mm]
\Gamma_\chi=2\cot\chi,\quad\Gamma_x=\i,\quad\Gamma_y=0\enspace.
\end{array}\qquad\right\}
\end{equation}
The momentum operators are
\begin{equation}
p_\chi=\hi\bigg(\Partial{\chi}+\cot\chi\bigg),\quad
p_x=\hi\bigg(\Partial{x}+\frac{\i}{2}\bigg),\quad
p_{y}=\hi\Partial{y}\enspace,
\label{pchi}
\end{equation}
and the Hamiltonian is given by
\begin{eqnarray}  
H&=&-\hbarm\Bigg[\Partialsq{\chi}+2\cot\chi\Partial{\chi}
+\frac{1}{\sin^2\chi}\bigg(\Partialsq{x}+{2\i x}\Partial{x}
+\e^{-2\i x}\Partialsq{y}\bigg)\Bigg]
\nonumber\\   &=&
    \frac{1}{2m}\Bigg[p_{\chi}^2
    +\frac{1}{\sin^2\chi}\bigg(p_{x}^2+\e^{-2\i x}p_{y}^2\Bigg)\Bigg]
    -\frac{\hbar^2}{8m}\bigg(4+\frac{1}{\sin^2\chi}\bigg)\enspace.
\end{eqnarray}  
For the path integral we find by separating off the $y$-path integration 
(plane waves):
\begin{eqnarray}
&&K^{(\SdreiC)}(\chi'',\chi',x'',x',y'',y';T)
\nonumber\\  &&
=\pathint{\chi}\pathint{x}\pathint{y}\e^{\i x}\sin^2\chi
\nonumber\\  &&\qquad\times
\exp\left\{\ih\int_0^T\left[\frac{m}{2}
\dot\chi^2+\sin^2\chi(\dot x^2+\e^{2\i x}\dot y^2)
+\frac{\hbar^2}{8m}\bigg(4+\frac{1}{\sin^2\chi}\bigg)\right]\dt\right\}
\qquad\qquad\qquad
\nonumber\\  &&
=(\sin\chi'\sin\chi'')^{-1/2}\e^{-\i(x''-x')/2}
\e^{\i\hbar T/2m}\int_{\bbbr}\frac{\e^{\i k_y(y''-y')}}{2\pi}
\nonumber\\  &&\qquad\times
\pathint{\chi}\pathint{x}\sin\chi
\nonumber\\  &&\qquad\times
\exp\left\{\ih\int_0^T\left[\frac{m}{2}
\dot\chi^2+\sin^2\chi\bigg(\dot x^2-\frac{\hbar^2k_y^2}{2m}\e^{2\i x}\bigg)
+\frac{\hbar^2}{8m\sin^2\chi}\right]\dt\right\}\enspace.
\end{eqnarray}
In the $x$-path integration we can use the result of the horicyclic
system (2) with energy spectrum $E_{n_x}=(n_x+\half)^2\hbar^2/2m$,
$n_x\in\bbbn_0$, yielding 
\begin{eqnarray}
&&K^{(\SdreiC)}(\chi'',\chi',x'',x',y'',y';T)
=(\sin\chi'\sin\chi'')^{-1}\e^{-\frac{\i}{2}(x''-x')}
\e^{\i\hbar T/2m}
\nonumber\\  &&\qquad\times
\int_{\bbbr}\frac{\e^{\i k_y(y''-y')}}{2\pi}
\sum_{n_x\in\bbbn_0} \bhalf
H^{(1)}_{n_x+1/2}(|k_y|\,\e^{-\i x''})H^{(1)}_{n_x+1/2}(|k_y|\,\e^{\i x'})
\nonumber\\  &&\qquad\times
\pathint{\chi}\exp\left[\ih\int_0^T\left(\frac{m}{2}
\dot\chi^2-\frac{\hbar^2}{2m}\frac{(n_x+\half)^2-\viert}{\sin^2\chi}
\right)\dt\right]
\nonumber\\  &&
=(\sin\chi'\sin\chi'')^{-1/2}\e^{-\frac{\i}{2}(x''-x')}
\e^{\i\hbar T/2m}\int_{\bbbr}\frac{\e^{\i k_y(y''-y')}}{2\pi}
\nonumber\\  &&\qquad\times
\sum_{n_x\in\bbbn_0} \bhalf
H^{(1)}_{n_x+1/2}(|k_y|\,\e^{-\i x''})H^{(1)}_{n_x+1/2}(|k_y|\,\e^{\i x'})
\nonumber\\  &&\qquad\times
\sum_{l\in\bbbn_0}\Bigg[(l+n_x+1)\frac{\Gamma(l+2n_x+2))}{l!}\Bigg]
P_{l+n_x+1/2}^{-n_x-1/2}(\cos\chi')P_{l+n_x+1/2}^{-n_x-1/2}(\cos\chi'')
\qquad\qquad\qquad
\nonumber\\  &&\qquad\times
\exp\bigg[-\ih\frac{\hbar^2(l+n_x)(l+n_x+2)}{2m}T\bigg]\enspace.
\end{eqnarray}
The path integral in the variable $\chi$ is just one for the
symmetric P\"oschl--Teller potential \cite{FLM,GRSh,KLEMUS}.
This yields for the wave-functions  on $\SdreiC$
\begin{eqnarray}
\Psi_{l,n_x,k_y}(\chi,x,y)&=&
\frac{\e^{\i k_y y}}{\sqrt{2\pi}}\cdot
\frac{1}{\sqrt{2}}\,\e^{-\i x/2}H^{(1)}_{n_x+1/2}(k_y\e^{-\i x})
\nonumber\\  &&\qquad\times
\Bigg[(l+n_x+1)\frac{\Gamma(l+2n_x+2))}{l!\sin\chi}\Bigg]^{1/2}
P_{l+n_x+1/2}^{-n_x-1/2}(\cos\chi)\enspace,\qquad
\end{eqnarray}
and the energy-spectrum (\ref{Energy-J}), by observing that we can
re-define $J=l+n_x$.  

\subsection{System 5: Horicyclic-Polar}
\message{System 5: Horicyclic-Polar}
This coordinate system is defined as
\begin{equation}
\left.\begin{array}{ll}
\displaystyle
 z_1=\half\Big[\e^{-\i x}+(1+\vrho^2)\e^{\i x}\Big]\qquad
&z_2=\i\rho\e^{\i x}\cos\vphi\\
\displaystyle
 z_3=\i\vrho\e^{\i x}\sin\vphi
&z_4=\frac{\i}{2}\Big[\e^{-\i x}-(1-\vrho^2)\e^{\i x}\Big]
\end{array}\quad\right\}
\end{equation}
$(\vphi\in[0,2\pi),x\in\bbbr\vrho>0)$. 
and the set of commuting operators is given by
\begin{equation}
\CL_1=(I_{42}^2+\i I_{21})^2+(I_{34}^2+\i I_{13})^2,\qquad 
\CL_2=I_{23}^2\enspace.
\end{equation}
For the metric terms we have 
\begin{equation}
\left.\begin{array}{l}
\d s^2=\d x^2+\e^{2\i x}(\d\vrho^2+\vrho^2\d\vphi^2)\enspace,
\\[2mm]
\sqrt{g}=\vrho\e^{2\i x}\enspace,
\\[2mm]
\Gamma_\vrho=\dfrac{1}{\vrho},\quad\Gamma_x=2\i,\quad\Gamma_\vphi=0\enspace,
\end{array}\qquad\right\}
\end{equation}
and the momentum operators are
\begin{equation}
p_x=\hi\bigg(\Partial{x}+\i\bigg),\quad
p_\vrho=\hi\bigg(\Partial{\vrho}+\frac{1}{2\vrho}\bigg),\quad
p_{\vphi}=\hi\Partial{\vphi}\enspace.
\end{equation}
Thus, the Hamiltonian is given by
\begin{eqnarray}  
H&=&-\hbarm\Bigg[\Partialsq{x}+2\i\Partial{x}
+\e^{-2\i x}\bigg(\Partialsq{\vrho}+\frac{1}{\vrho}\Partial{\vrho}
+\frac{1}{\vrho^2}\Partialsq{\vphi}\bigg)\Bigg]
\nonumber\\   &=&
    \frac{1}{2m}\Bigg[p_{x}^2
    +\e^{-2\i x}\bigg(p_{\vrho}^2+\frac{1}{\vrho^2}p_\vphi^2\Bigg)\Bigg]
    -\e^{-2\i x}\frac{\hbar^2}{8m\vrho^2}\enspace.
\end{eqnarray}  
In the following path integral, the variable $\vphi$ can be separated off
in terms of circular wave, the variable $\vrho$ in terms of a radial path integral
(free motion), and for the remaining path integration in $x$
we find similarly as in the horicyclic system (2)
\begin{eqnarray}
&&K^{(\SdreiC)}(x'',x',\vrho'',\vrho',\vphi'',\vphi';T)
\nonumber\\  &&
=\pathint{x}\pathint{\vrho}\pathint{\vphi}\vrho\e^{2\i x}
\nonumber\\  &&\qquad\times
\exp\left\{\ih\int_0^T\left[\frac{m}{2}
\dot x^2+\e^{2\i x}(\dot\vrho^2+\vrho^2\dot\vphi^2)
+\e^{-2\i x}\frac{\hbar^2}{8m\vrho^2}\right]\dt\right\}
\nonumber\\  &&
=e^{-\i (x''-x')}\sum_{\nu\in\bbbz}\frac{\e^{\i\nu(\vphi''-\vphi')}}{2\pi}
\int_0^\infty \d k_\vrho k_\vrho J_\nu(k_\vrho\vrho')J_\nu(k_\vrho\vrho'')
\nonumber\\  &&\qquad\times
\pathint{x}\exp\left[\ih\int_0^T\left(\frac{m}{2}
\dot x^2-\e^{-2\i x}\frac{\hbar^2k_\vrho^2}{2m}\right)\dt\right]
\qquad\qquad\qquad\qquad\qquad\qquad
\nonumber\\  &&
=e^{-\i (x''-x')}\sum_{\nu\in\bbbz}\frac{\e^{\i\nu(\vphi''-\vphi')}}{2\pi}
\int_0^\infty \d k_\vrho k_\vrho J_\nu(k_\vrho\vrho')J_\nu(k_\vrho\vrho'')
\nonumber\\  &&\qquad\times
\sum_{J\in\bbbn_0} \bhalf
H^{(1)}_{J+1/2}\big(|k_\vrho|\,\e^{-\i x''}\big)
H^{(1)}_{J+1/2}\big(|k_\vrho|\,\e^{\i x'}\big)
\exp\Bigg[-\ih\frac{\hbar^2J(J+2)}{2m}T\Bigg]\enspace.
\end{eqnarray}
The wave-functions on $\SdreiC$  are given by ($k_\vrho>0)$
\begin{equation}
\Psi_{Jk_\vrho\nu}(x,\vrho,\vphi)=
\frac{\e^{\i\nu\vphi}}{\sqrt{2\pi}}
 \sqrt{k_\vrho}\, J_\nu(k_\vrho\vrho')\e^{-\i x}\cdot\frac{1}{\sqrt{2}}
H^{(1)}_{J+1/2}(k_\vrho\,\e^{-\i x})\enspace,
\end{equation}
with the energy-spectrum (\ref{Energy-J}). 

\subsection{System 6: Sphero-Elliptic}
\message{System 6: Sphero-Elliptic}
This coordinate system is defined as
\begin{equation}
\left.\begin{array}{ll}
\displaystyle
 z_1=\sin\chi\sn(\alpha,k)\dn(\beta,k')\qquad
&z_2=\sin\chi\cn(\alpha,k)\cn(\beta,k')
 \\[3mm]
\displaystyle
 z_3=\sin\chi\dn(\alpha,k)\sn(\beta,k')
&z_4=\cos\chi.
\end{array}\quad\right\}
\end{equation}
$(\chi\in(0,\pi),\alpha\in[-K,K],\beta\in[-2K',2K'])$.
The set of commuting operators is given by
\begin{equation}
\CL_1=I_{12}^2+I_{13}^2+I_{23}^2,\qquad 
\CL_2=I_{23}^2+k^2I_{13}^2\enspace.
\end{equation}
The elliptic coordinate system reads in algebraic form as follows
($a_1\leq\vrho_1\leq a_2\leq\vrho_2\leq a_3$)
\begin{equation}
  s_1^2=R^2\dfrac{(\vrho_1-a_1)(\vrho_2-a_1)}{(a_2-a_1)(a_3-a_1)},\
  s_2^2=R^2\dfrac{(\vrho_1-a_2)(\vrho_2-a_2)}{(a_3-a_2)(a_1-a_2)},\
  s_3^2=R^2\dfrac{(\vrho_1-a_3)(\vrho_2-a_3)}{(a_1-a_3)(a_2-a_3)}
  \enspace.
\end{equation}
If we put $\vrho_1=a_1+(a_2-a_1)\sn^2(\alpha,k)$ and
$\vrho_2=a_2+(a_3-a_2)\cn^2(\beta,k')$, where $\sn(\alpha,k),
\cn(\alpha,k)$ and $\dn(\alpha,k)$ are the Jacobi elliptic functions
with modulus $k$, we obtain for the coordinates $\vec s$ on the sphere
(also called conical coordinates)
\begin{equation}\begin{array}{ll}
  s_1=R\sn(\alpha,k)\dn(\beta,k')\qquad
     &(-K\leq \alpha\leq K)\enspace,   \\[2mm]
  s_2=R\cn(\alpha,k)\cn(\beta,k')
     &(-2K'\leq\beta\leq2K')\enspace, \\[2mm]
  s_3=R\dn(\alpha,k)\sn(\beta,k')\enspace, &    \end{array}
\end{equation}
where
\begin{equation}
  k^2={a_2-a_1\over a_3-a_1}=\sin^2f\enspace,\qquad
  {k'}^2={a_3-a_2\over a_3-a_1}=\cos^2f\enspace,\qquad
  k^2+{k'}^2=1\enspace.
\end{equation}
$K=K(k)={\pi\over2}{_2}F_1(\half,\half;1;k^2)$ and $K'=K(k')$ are
complete elliptic integrals, and $2f$ is the interfocus distance on
the upper semi-sphere of the ellipses on the sphere. Note the relations
$\cn^2\alpha+\sn^2\alpha=1$ and $\dn^2\alpha=1-k^2\sn^2\alpha$. In the
following we omit the moduli $k$ and $k'$ of the Jacobi elliptic
functions if it is obvious that the variable $\alpha$ goes with $k$ and
$\beta$ goes with $k'$. 
The metric terms are now
\begin{equation}
\left.\begin{array}{l}
\d s^2=\d\chi^2+\sin^2\chi(k^2\cn^2\alpha+{k'}^2\cn^2\beta)
                          (\d\alpha^2+\d\beta^2)\enspace,
\\[2mm]
\sqrt{g}=(k^2\cn^2\alpha+{k'}^2\cn^2\beta)\sin^2\chi\enspace,
\\
\Gamma_\chi=2\cot\chi\enspace,\quad
\Gamma_\alpha=-2\dfrac{k^2\sn\alpha\cn\alpha\dn\alpha}{
  k^2\cn^2\alpha+{k'}^2\cn^2\alpha},\quad
\Gamma_\beta=-2\dfrac{{k'}^2\sn\beta\cn\beta\dn\beta}{
  k^2\cn^2\beta+{k'}^2\cn^2\beta}\enspace.
\end{array}\quad\right\}
\end{equation}
The momentum operators are ($p_\chi$ as in (\ref{pchi}))
\begin{equation}
  p_\alpha=\hi\bigg({\partial\over\partial\alpha}
         -{k^2\sn\alpha\cn\alpha\dn\alpha\over
  k^2\cn^2\alpha+{k'}^2\cn^2\alpha}\bigg)\enspace,\quad
  p_\beta=\hi\bigg({\partial\over\partial\beta}
         -{{k'}^2\sn\beta\cn\beta\dn\beta\over
  k^2\cn^2\beta+{k'}^2\cn^2\beta}\bigg)\enspace,
\end{equation}
and the Hamiltonian has the form
\begin{eqnarray} 
  H&&=-\hbarm\Bigg[\Partialsq{\chi}+2\cot\chi\Partial{\chi}
   +\frac{1}{\sin^2\chi}{1\over k^2\cn^2\alpha+{k'}^2\cn^2\alpha}\bigg(
   {\partial^2\over\partial\alpha^2}+{\partial^2\over\partial\beta^2}
  \bigg)\Bigg]
\nonumber\\   &&
  ={1\over2m}\left[p_\chi^2+\frac{1}{\sin^2\chi}
   {1\over\sqrt{k^2\cn^2\alpha+{k'}^2\cn^2\beta}}
   (p_\alpha^2+p_\beta^2){1\over\sqrt{k^2\cn^2\alpha+{k'}^2\cn^2\beta}}
   \right]-\hbarm\enspace.\qquad
\end{eqnarray} 
The corresponding path integral representation has been discussed in 
\cite{GROad,GROPOb} and we just state the result:
\begin{eqnarray}       & &\!\!\!\!\!\!\!
K^{(\SdreiC)}(\chi'',\chi',\alpha'',\alpha',\beta'',\beta';T)
         \nonumber\\   & &\!\!\!\!\!\!\!
 =\pathint\chi\sin^2\chi\pathint\alpha\pathint\beta\Emetric
         \nonumber\\   & &\!\!\!\!\!\!\!\qquad\qquad\times
  \exp\Bigg\{{\i m\over2\hbar}\intt\Big[\dot\chi^2+\sin^2\chi
   \Emetric(\dot\alpha^2+\dot\beta^2)\Big]dt+{\i\hbar T\over2m}\Bigg\}
         \nonumber\\   & &\!\!\!\!\!\!\!
 =\sum_{J=0}^\infty\sum_{l=-J}^J\sum_{\lambda}\sum_{p,q=\pm}
     \Lambda_{l,h}^p(\alpha'')\Lambda_{l,h}^{p\,*}(\alpha')
     \Lambda_{l,\tilde h}^q(\beta'')\Lambda_{l,\tilde h}^{q\,*}(\beta')
         \nonumber\\   & &\!\!\!\!\!\!\!\qquad\qquad\times
  (J+1){(l+J+1)!\over|J-l|!}\,\e^{-\i\hbar TJ(J+2)/2m}
  P_{J+1/2}^{-l-1/2}(\sin\chi'')P_{J+1/2}^{-l-1/2}(\sin\chi')
\enspace.\qquad\qquad
\end{eqnarray}
The solution in the variable $\chi$ is again of the symmetric P\"oschl--Teller type.
For the {\it periodic Lam\'e polynomials} $\Lambda_{ln}^p(z)$ we have adopted the
notation of \cite{PAWIa}. In \cite{PAWIa} it 
is shown that the wave-functions of the spherical basis $|lm\!>$ can be
expanded into the wave-functions of the elliptical basis $|l\lambda\!>$
and vice versa. The are Lam\'e polynomials $\Lambda_{lh}^p$ are satisfying
($k^2=1-{k'}^2$) 
\begin{equation}\left.\begin{array}{l}
  \dfrac{\d^2\Lambda_{lh}^p}{\d\alpha^2}
  +\big[h-l(l+1)k^2\sn^2(\alpha,k)\big]\Lambda_{lh}^p=0\enspace,  \\
  \Lambda_{lh}^p(-\alpha)=p\Lambda_{lh}(\alpha)\enspace,\qquad
  h=-\dfrac{\lambda}{4}+l(l+1)\enspace.
  \end{array}\qquad\right\}
\end{equation}
The functions $\Lambda_{lh'}^q(\beta)$ satisfies the same equation with
$\alpha\to\beta,k\to k',h\to\tilde h=\lambda/4$ and $p\to q$. These
functions are also called {\it ellipsoidal harmonics} which satisfy the
orthonormality relation 
\begin{equation}
  \int_{-K}^Kd\alpha\int_{-2K'}^{2K'}d\beta\Emetric
     \Lambda_{l',\tilde h}^{p'}(\alpha)\Lambda_{l,h}^{p\,*}(\alpha)
     \Lambda_{l',\tilde h'}^{q'}(\beta)\Lambda_{l,h'}^{q\,*}(\beta)
  =\delta_{ll'}\delta_{qq'}\delta_{pp'}\delta_{h\tilde h}\enspace.
\end{equation}
Here $\lambda$ is the eigenvalue of the operator $E=\CL_2$ \cite{PAWIa} 
which commutes with the Hamiltonian. 
\newline
The wave-functions on $\SdreiC$  have the form
\begin{equation}
\Psi_{J,l,\lambda,p}=\Lambda_{l,h}^p(\alpha)\Lambda_{l,\tilde h}^q(\beta)
\sqrt{(J+1){(l+J+1)!\over|J-l|!}}\,P_{J+1/2}^{-l-1/2}(\sin\chi)\enspace,
\end{equation}
and the energy-spectrum (\ref{Energy-J}). 
The sphero-elliptic system exists on the three-dimensional sphere (System II.)
and on the three-dimensional hyperboloid (System III.).

\subsection{System 7: Spherical-degenerate elliptic I}
\message{System 7: Spherical-degenerate elliptic I}
This coordinate system is defined as
\begin{equation}
\left.\begin{array}{ll}
\displaystyle
 z_1=\half\sin\chi\bigg(\frac{\cosh\tau_2}{\cosh\tau_1}
                       +\frac{\cosh\tau_1}{\cosh\tau_2}\bigg)
&z_2=\sin\chi\tanh\tau_1\tanh\tau_2\\[3mm]
\displaystyle
 z_3=\sin\chi\bigg[\frac{1}{\cosh\tau_1\cosh\tau_2}
      -\half\bigg(\frac{\cosh\tau_2}{\cosh\tau_1}
                 +\frac{\cosh\tau_1}{\cosh\tau_2}\bigg)\bigg]\qquad
&z_4=\cos\chi
\end{array}\quad\right\}
\end{equation}
$(\chi\in[0,\pi),\tau_1,\tau_2\in\bbbr)$.
The set of commuting operators is given by
\begin{equation}
\CL_1=I_{12}^2+I_{13}^2+I_{23}^2,\qquad 
\CL_2=-I_{12}^2+I_{23}^2+\i\{I_{31},I_{32}\}\enspace.
\end{equation}
$\{I,J\}=IJ+JI$ denotes the anti-commutator of the operators $I$ and $J$.
The metric terms are 
\begin{equation}
\left.\begin{array}{l}
\displaystyle
\d s^2=\d\chi^2+\sin^2\chi\bigg(\frac{1}{\cosh^2\tau_1}
      -\frac{1}{\cosh^2\tau_2}\bigg)(\d\tau_1^2-\d\tau_2^2)\enspace,
\\[3mm]
\displaystyle
\sqrt{g}=\sin^2\chi\bigg(\frac{1}{\cosh^2\tau_1}
      -\frac{1}{\cosh^2\tau_2}\bigg)\enspace,
\\[3mm]
\Gamma_\chi=2\cot\chi,\quad
\Gamma_{\tau_{1,2}}=\mp2\dfrac{\sinh\tau_{1,2}}{\cosh^3\tau_{1,2}}
\cdot\dfrac{1}{\cosh^{-2}\tau_1-\cosh^{-2}\tau_2}
\enspace.
\end{array}\quad\right\}
\end{equation}
The momentum operators have the form
\begin{equation}
p_\chi=\hi\bigg(\Partial{\chi}+\cot\chi\bigg),\quad
p_{\tau_1}=\hi\bigg(\Partial{\tau_1}
              +\bhalf\Gamma_{\tau_1}\bigg),\quad
p_{\tau_2}=\hi\bigg(\Partial{\tau_2}
              +\bhalf\Gamma_{\tau_2}\bigg)\enspace,
\end{equation}
and the Hamiltonian reads
\begin{eqnarray} 
  H&=&-\hbarm\Bigg[\Partialsq{\chi}+2\cot\chi\Partial{\chi}
\nonumber\\  &&\qquad\qquad\qquad
   +\frac{1}{\sin^2\chi}\bigg(\frac{1}{\cosh^2\tau_1}
      -\frac{1}{\cosh^2\tau_2}\bigg)^{-1}\bigg(
   \Partialsq{\tau_1}+\bhalf\Gamma_{\tau_1}\Partial{\tau_1}
  -\Partialsq{\tau_2}-\bhalf\Gamma_{\tau_2}\Partial{\tau_2}\bigg)\Bigg]
\nonumber\\   &=&
  {1\over2m}\left[p_\chi^2+\frac{1}{\sin^2\chi}
  \bigg(\frac{1}{\cosh^2\tau_1}
      -\frac{1}{\cosh^2\tau_2}\bigg)^{-1/2}
   (p_{\tau_1}^2-p_{\tau_2}^2)
   \bigg(\frac{1}{\cosh^2\tau_1}
      -\frac{1}{\cosh^2\tau_2}\bigg)^{-1/2}\,\right]
\nonumber\\  &&\qquad\qquad\qquad
   -\hbaram\bigg(4+{1\over\sin^2\chi}\bigg)\enspace.
\end{eqnarray} 
In the calculation I use now for the ($\tau_1,\tau_2$)-subpath integration 
a path integral solution  on the two-dimensional hyperboloid 
\cite[p.97]{GROad}, i.e. the kernel in terms of elliptic parabolic coordinates.
The $\chi$-path integration is the usual symmetric P\"oschl--Teller case
\cite{FLM,GRSh,KLEMUS}, therefore we obtain
\begin{eqnarray}
&&K^{(\SdreiC)}(\chi'',\chi',\tau_1'',\tau_1',\tau_2'',\tau_2';T)
\nonumber\\  &&
=\pathint{\chi}\pathint{\tau_1}\pathint{\tau_2}
\sin^2\chi\bigg(\frac{1}{\cosh^2\tau_1}
      -\frac{1}{\cosh^2\tau_2}\bigg)
\nonumber\\  &&\qquad\times
\exp\left\{\ih\int_0^T\left[\frac{m}{2}
\dot\chi^2+\sin^2\chi\bigg(\frac{1}{\cosh^2\tau_1}
      -\frac{1}{\cosh^2\tau_2}\bigg)(\dot\tau_1^2-\dot\tau_2^2)
+\hbaram\bigg(4+{1\over\sin^2\chi}\bigg)\right]\dt\right\}
\nonumber\\  &&
=(\sin\chi'\sin\chi'')^{-1}\e^{\i\hbar T/2m}
\int_0^\infty dp\,p\sinh\pi p
  \int_0^\infty{dk\,k\sinh\pi k\over(\cosh^2\pi k+\sinh^2\pi p)^2}\,
\nonumber\\  &&\qquad\times
  \sum_{\epsilon,\epsilon'=\pm1}
  P^{ \i k}_{\i p-1/2}(\epsilon \tanh\tau_1'')
  P^{-\i k}_{\i p-1/2}(\epsilon \tanh\tau_1' )
  P^{ \i p}_{\i k-1/2}(\epsilon'\tanh\tau_2'')
  P^{-\i p}_{\i k-1/2}(\epsilon'\tanh\tau_2' )
\nonumber\\  &&\qquad\times
\pathint{\chi}\exp\left[\ih\int_0^T\left(\frac{m}{2}
\dot\chi^2-\frac{\hbar^2}{2m}\frac{p^2-\viert}{\sin^2\chi}\right)\dt\right]
\nonumber\\  &&
=(\sin\chi'\sin\chi'')^{-1}\e^{\i\hbar T/2m}
\int_0^\infty dp\,p\sinh\pi p
\int_0^\infty{dk\,k\sinh\pi k\over(\cosh^2\pi k+\sinh^2\pi p)^2}\,
\nonumber\\  &&\qquad\times
  \sum_{\epsilon,\epsilon'=\pm1}
  P^{ \i k}_{\i p-1/2}(\epsilon \tanh\tau_1'')
  P^{-\i k}_{\i p-1/2}(\epsilon \tanh\tau_1' )
  P^{ \i p}_{\i k-1/2}(\epsilon'\tanh\tau_2'')
  P^{-\i p}_{\i k-1/2}(\epsilon'\tanh\tau_2' )
\nonumber\\  &&\qquad\times
\sum_{J\in\bbbn_0}\Bigg[(J+p+\bhalf)\frac{\Gamma(p+2J+1)}{J!}\Bigg]
P_{J+p+1/2}^{-J-1/2}(\cos\chi')P_{J+p+1/2}^{-J-1/2}(\cos\chi'')
\nonumber\\  &&\qquad\times
\exp\bigg[-\ih\frac{\hbar^2J(J+2)}{2m}T\bigg]\enspace.
\end{eqnarray}
The wave-functions on $\SdreiC$  have the form
\begin{eqnarray}
&&\Psi_{J,p,k,\pm}(\chi,\tau_1,\tau_2)
=\frac{\sqrt{p\sinh\pi pk\sinh\pi k}}{\cosh^2\pi k+\sinh^2\pi p}
  P^{ \i k}_{\i p-1/2}(\epsilon \tanh\tau_1)
  P^{ \i p}_{\i k-1/2}(\epsilon'\tanh\tau_2)
\nonumber\\  &&\qquad\times
(\sin\chi)^{-1}\Bigg[(J+p+\bhalf)\frac{\Gamma(p+2J+1)}{J!}\Bigg]^{1/2}
P_{J+p+1/2}^{-J-1/2}(\cos\chi')\enspace,
\end{eqnarray}
and the energy-spectrum (\ref{Energy-J}). 

\subsection{System 8: Spherical-degenerate elliptic II}
\message{System 8: Spherical-degenerate elliptic II}
This coordinate system is defined as
\begin{equation}
\left.\begin{array}{ll}
\displaystyle
 z_1=-\frac{\i\sin\chi}{8\xi\eta}\Big[(\xi^2-\eta^2)^2+4\big]\qquad
&\displaystyle
z_2=\frac{\sin\chi}{2\xi\eta}(\xi^2+\eta^2)^2
\\[3mm]
\displaystyle
 z_3=\frac{\sin\chi}{8\xi\eta}\Big[-(\xi^2-\eta^2)^2+4\big]              
&\displaystyle
z_4=\cos\chi-\{I_{12},I_{13}\}+\i\{I_{12},I_{23}\}
\end{array}\quad\right\}\end{equation}
$(\chi\in(0,\pi),\xi,\eta>0)$. The set of commuting operators is given by
\begin{equation}
\CL_1=I_{12}^2+I_{13}^2+I_{23}^2,\qquad 
\CL_2=-I_{12}^2+I_{13}^2+\i\{I_{12},I_{23}\}\enspace,
\end{equation}
and the metric terms are
\begin{equation}
\left.\begin{array}{l}
\displaystyle
\d s^2=\d\chi^2+\sin^2\chi\bigg(\frac{1}{\eta^2}-\frac{1}{\xi^2}\bigg)
       (\d\xi^2-\d\eta^2)\enspace,
\\[3mm]
\displaystyle
\sqrt{g}=\sin^2\chi\bigg(\frac{1}{\eta^2}-\frac{1}{\xi^2}\bigg)\enspace,
\\
\Gamma_\chi=2\cot\chi,\quad
\Gamma_\xi =-\dfrac{2}{\xi^3}\cdot\dfrac{1}{1/\eta^2-1/\xi^2},\quad
\Gamma_\eta= \dfrac{2}{\eta^3}\cdot\dfrac{1}{1/\eta^2-1/\xi^2},\quad
\enspace.
\end{array}\quad\right\}
\end{equation}
The momentum operators have the form
\begin{equation}
p_\chi=\hi\bigg(\Partial{\chi}+\cot\chi\bigg),\quad
p_\xi =\hi\bigg(\Partial{\xi}+\bhalf\Gamma_\xi\bigg),\quad
p_\eta=\hi\bigg(\Partial{\eta}+\bhalf\Gamma_\eta\bigg)\enspace,
\end{equation}
and the Hamiltonian reads
\begin{eqnarray} 
  H&=&-\hbarm\Bigg[\Partialsq{\chi}+2\cot\chi\Partial{\chi}
   +\frac{1}{\sin^2\chi}
\bigg(\frac{1}{\eta^2}-\frac{1}{\xi^2}\bigg)^{-1}\bigg(
   \Partialsq{\xi}+\bhalf\Gamma_{\xi}\Partial{\xi}
  -\Partialsq{\eta}+\bhalf\Gamma_{\eta}\Partial{\eta}\bigg)\Bigg]
\nonumber\\   &=&
  {1\over2m}\left[p_\chi^2+\frac{1}{\sin^2\chi}
  \bigg(\frac{1}{\eta^2}-\frac{1}{\xi^2}\bigg)^{-1/2}
   (p_{\xi}^2-p_{\eta}^2)
   \bigg(\frac{1}{\eta^2}-\frac{1}{\xi^2}\bigg)^{-1/2}\,\right]
   -\hbaram\bigg(4+{1\over\sin^2\chi}\bigg)\enspace.
\nonumber\\  &&
\end{eqnarray} 
For the path integral we obtain
\begin{eqnarray}
&&K^{(\SdreiC)}(\chi'',\chi',\xi'',\xi',\eta'',\eta';T)
=\pathint{\chi}\pathint{\xi}\pathint{\eta}
\sin^2\chi\bigg(\frac{1}{\eta^2}-\frac{1}{\xi^2}\bigg)
\nonumber\\  &&\qquad\times
\exp\left\{\ih\int_0^T\left[\frac{m}{2}\dot\chi^2
+\sin^2\chi\bigg(\frac{1}{\eta^2}-\frac{1}{\xi^2}\bigg)(\dot\xi^2-\dot\eta^2)
+\hbaram\bigg(4+{1\over\sin^2\chi}\bigg)\right]\dt\right\}\enspace.\qquad
\label{Pathintegral-VIII}
\end{eqnarray}
For the evaluation of this path integral we first consider the
$(\xi,\eta)$-subpath integration, denoted by $\hat K(\xi'',\xi',\eta'',\eta';T)$. 
By the usual technique of space-time transformation we obtain for the
corresponding transformed path integral
\begin{equation}
\hat K(\xi'',\xi',\eta'',\eta';s'')
=\pathints{\xi}\pathints{\eta}
\exp\left\{\ih\int_0^{s''}\Bigg[\frac{m}{2}(\dot\xi^2-\dot\eta^2)
-\frac{\CE}{\xi^2}+\frac{\CE}{\eta^2}\Bigg]\d s\right\}.
\end{equation}
We therefore obtain two path integrations for the inverse-square
potential. Problems like these have been discussed in \cite{GROad,GROas}.
The corresponding Green function we can write as follows:
\begin{eqnarray}
&&\hat G(\xi'',\xi',\eta'',\eta';\CE)
=\int_0^\infty\d s''\hat K(\xi'',\xi',\eta'',\eta';s'')
\nonumber\\   &&
=\frac{4m^2}{\hbar^3}\sqrt{\xi'\xi''\eta'\eta''}\int\frac{\d\CE}{2\pi\i}
I_{\lambda}\bigg(\sqrt{2m\CE}\,\frac{\xi_>}{\hbar}\bigg)
K_{\lambda}\bigg(\sqrt{2m\CE}\,\frac{\xi_>}{\hbar}\bigg)
I_{\tilde\lambda}\bigg(\sqrt{-2m\CE}\,\frac{\eta_<}{\hbar}\bigg)
K_{\tilde\lambda}\bigg(\sqrt{-2m\CE}\,\frac{\eta_>}{\hbar}\bigg)
\nonumber\\   &&
=\frac{m^2}{\hbar^3}\sqrt{\xi'\xi''\eta'\eta''}
\int_0^\infty\frac{\d s''}{s''}\int\frac{\d\CE}{2\pi\i}\,\e^{\i\CE s''/\hbar}
\nonumber\\   &&\qquad\times
\exp\bigg[\frac{m}{2\i\hbar s''}({\xi'}^2+{\xi''}^2)\bigg]
I_\lambda\bigg(\frac{\i m\xi'\xi''}{\hbar s''}\bigg)
I_{\tilde\lambda}\bigg(\sqrt{-2m\CE}\,\frac{\eta_<}{\hbar}\bigg)
K_{\tilde\lambda}\bigg(\sqrt{-2m\CE}\,\frac{\eta_>}{\hbar}\bigg)
\end{eqnarray}
($\CE=(\lambda^2-\viert)/2m$). 
$\xi_{<,>}$ denotes the smaller/larger of $\xi',\xi''$, and similarly for
$\eta$. By means of a similar analysis as in 
\cite{GROad,GROas} the Green function $\hat G(\CE)$ is found to read
\begin{eqnarray}
&&\hat G(\xi'',\xi',\eta'',\eta';\CE)
=\frac{1}{4\pi^2}\sqrt{\mu'\mu''\nu'\nu''}
\int\d\kappa\int_0^\infty\frac{\d k\,k\sinh^2\pi k}
{\hbar^2k^2/2m-\CE}
\nonumber\\   &&\qquad\qquad\qquad\qquad\times
H_{-\i\tilde p}^{(1)}\big(\sqrt{\kappa}\,\xi'\big)
H_{-\i\tilde p}^{(1)*}\big(\sqrt{\kappa}\,\xi''\big)
H_{-\i\tilde p}^{(1)}\big(\sqrt{\kappa}\,\eta'\big)
H_{-\i\tilde p}^{(1)*}\big(\sqrt{\kappa}\,\eta''\big)\enspace.
\end{eqnarray}
We insert this result in (\ref{Pathintegral-VIII}) and get together with 
the remaining path integral in $\chi$ (symmetric P\"oschl--Teller case):
\begin{eqnarray}
&&K^{(\SdreiC)}(\chi'',\chi',\xi'',\xi',\eta'',\eta';T)
\nonumber\\   &&
=\frac{1}{4\pi^2}\sqrt{\mu'\mu''\nu'\nu''}
\int\d k\int_0^\infty \d p\,p\sinh^2\pi p
H_p^{(1)}\big(\sqrt{k}\,\xi'\big)
H_p^{(1)*}\big(\sqrt{k}\,\xi''\big)
H_p^{(1)}\big(\sqrt{k}\,\eta'\big)
H_p^{(1)*}\big(\sqrt{k}\,\eta''\big)
\nonumber\\  &&\qquad\times
\sum_{J\in\bbbn_0}\Bigg[(J+p+\bhalf)\frac{\Gamma(p+2J+1)}{J!}\Bigg]
P_{J+p+1/2}^{-J-1/2}(\cos\chi')P_{J+p+1/2}^{-J-1/2}(\cos\chi'')
\nonumber\\  &&\qquad\times
\exp\bigg[-\ih\frac{\hbar^2J(J+2)}{2m}T\bigg]\enspace.
\end{eqnarray}
The wave-functions on $\SdreiC$  have the form
\begin{eqnarray}
&&\Psi_{J,k,\kappa}(\chi,\xi,\eta)
=\frac{\sqrt{p\mu\nu}}{2\pi}\sinh\pi p
H_p^{(1)}\big(\sqrt{k}\,\xi\big)H_p^{(1)}\big(\sqrt{k}\,\eta\big)
\nonumber\\  &&\qquad\times
(\sin\chi)^{-1}\Bigg[(J+p+\bhalf)\frac{\Gamma(p+2J+1)}{J!}\Bigg]^{1/2}
P_{J+p+1/2}^{-J-1/2}(\cos\chi')\enspace,
\end{eqnarray}
and the energy-spectrum (\ref{Energy-J}). 

\subsection{System 9: Horicyclic-Elliptic}
\message{System 9: Horicyclic-Elliptic}
This coordinate system is defined as
\begin{equation}
\left.\begin{array}{ll}
\displaystyle
 z_1=\half\Big[\e^{-\i x}+(1+\cosh^2\tau_1+\sinh^2\tau_2)\e^{\i x}\Big]
&\displaystyle
z_2=\i\cosh\tau_1\cosh\tau_2\e^{\i x}
\\[3mm]
\displaystyle
 z_3=\sinh\tau_1\sinh\tau_2\e^{\i x}
&\displaystyle
z_4=\frac{\i}{2}
     \Big[\e^{-\i x}+(-1+\cosh^2\tau_1+\sinh^2\tau_2)\e^{\i x}\Big]
\end{array}\ \right\}
\end{equation}
$(x,\tau_1,\tau_2\in\bbbr)$,
the set of commuting operators read
\begin{equation}
\CL_1=(I_{42}+\i I_{21})^2+(I_{34}+\i I_{13})^2,\qquad
\CL_2=I_{23}^2+(I_{34}+\i I_{13})^2\enspace,
\end{equation}
the metric terms are given by
\begin{equation}
\left.\begin{array}{l}
\d s^2=\d x+\e^{2\i x}(\cosh^2\tau_1-\cosh^2\tau_2)(\d\tau_1^2-\d\tau_2^2)
\enspace,
\\[3mm]
\sqrt{g}=\e^{2\i x}(\cosh^2\tau_1-\cosh^2\tau_2)\enspace,
\\[3mm]
\Gamma_x=2\i,\quad
\Gamma_{\tau_1}=\dfrac{2\sinh\tau_2\cosh\tau_2}{\cosh^2\tau_2-\cosh^2\tau_2},
\quad
\Gamma_{\tau_1}=\dfrac{-2\sinh\tau_2\cosh\tau_2}{\cosh^2\tau_2-\cosh^2\tau_2}
\enspace,
\end{array}\quad\right\}
\end{equation}
the momentum operators have the form
\begin{equation}
p_x=\hi\bigg(\Partial{x}+\i\bigg),\quad
p_{\tau_1}=\hi\bigg(\Partial{\tau_1}+\bhalf\Gamma_{\tau_1}\bigg),\quad
p_{\tau_2}=\hi\bigg(\Partial{\tau_2}+\bhalf\Gamma_{\tau_2}\bigg)\enspace,
\end{equation}
and the Hamiltonian reads
\begin{eqnarray} 
  H&=&-\hbarm\Bigg[\Partialsq{x}+2\i\Partial{\chi}
   +\frac{\e^{-2\i x}}{\cosh^2\tau_2-\cosh^2\tau_2}
   \Bigg(\Partialsq{\tau_1}+\Gamma_{\tau_1}\Partial{\tau_1}
  -\Partialsq{\tau_2}-\Gamma_{\tau_2}\Partial{\tau_2}\Bigg)\Bigg]\qquad
\nonumber\\   &=&
  {1\over2m}\left[p_x^2+\frac{\e^{-2\i x}}{\sqrt{\cosh^2\tau_2-\cosh^2\tau_2}}
   (p_{\tau_1}^2-p_{\tau_2}^2)
   \frac{1}{\sqrt{\cosh^2\tau_2-\cosh^2\tau_2}}\right]-\hbarm\enspace.
\end{eqnarray} 
For the path integral representation we obtain
\begin{eqnarray}
&&K^{(\SdreiC)}(x'',x',\tau_1'',\tau_1',\tau_2'',\tau_2';T)
\nonumber\\  &&
=\pathint{x}\pathint{\tau_1}\pathint{\tau_2}
\e^{2\i x}(\cosh^2\tau_1-\cosh^2\tau_2)\qquad\qquad\qquad\qquad
\nonumber\\  &&\qquad\times
\exp\left\{\ih\int_0^T\Bigg[\frac{m}{2}\dot x^2
+\e^{2\i x}(\cosh^2\tau_1-\cosh^2\tau_2)(\dot\tau_1^2-\dot\tau_2^2)
+\frac{\hbar^2}{2m}\Bigg]\dt\right\}\enspace.
\end{eqnarray}
We first consider the path integral in $(\tau_1,\tau_2)$. It has exactly 
the form of the path integral representation of an elliptic coordinate 
system on the two-dimensional pseudo-Euclidean plane \cite{GROad,KALc,KAMIo}.
We can immediately use the corresponding result, and together with the
$x$-path integration of the horicyclic-polar system this yields the final result
\begin{eqnarray}
&&K^{(\SdreiC)}(x'',x',\tau_1'',\tau_1',\tau_2'',\tau_2';T)
\nonumber\\  &&
 ={1\over8\pi}\int_0^\infty pdp\int_{\bbbr}dk\,\e^{-\pi k}
   \Me_{\i k}(\tau_2'';\hhbox{p^2\over4})
   \Me_{\i k}^*(\tau_2';\hhbox{p^2\over4})
   M_{\i k}^{(3)}(\tau_1'';\hhbox{p\over2})
   M_{\i k}^{(3)\,*}(\tau_1';\hhbox{p\over2})\qquad\qquad
\nonumber\\  &&\qquad\times
\sum_{J\in\bbbn_0} \bhalf
H^{(1)}_{J+1/2}\big(|k_\vrho|\,\e^{-\i x''}\big)
H^{(1)}_{J+1/2}\big(|k_\vrho|\,\e^{\i x'}\big)
\exp\Bigg[-\ih\frac{\hbar^2J(J+2)}{2m}T\Bigg]\enspace.
\end{eqnarray}
The wave-functions on $\SdreiC$  are given by
\begin{equation}
\Psi_{Jpk}(x,\tau_1,\tau_2)=
 \sqrt{p\over16\pi} \e^{-\pi k/2}
   \Me_{\i k}(\tau_2;\hhbox{p^2\over4})
   M_{\i k}^{(3)}(\tau_1;\hhbox{p\over2})
\e^{-\i x} H^{(1)}_{J+1/2}(p\,\e^{-\i x})\enspace,
\end{equation}
and the energy-spectrum (\ref{Energy-J}). 
$\Me_\nu(z)$ and $\Me_\nu^{(3)}(z)$ are Mathieu-functions \cite{MESCH}, which are
typical for the quantum motion in elliptic coordinates in two
dimensions \cite{GROab}.

\subsection{System 10: Horicyclic-Hyperbolic}
\message{System 10: Horicyclic-Parabolic I}
This coordinate system is defined as
\begin{equation}
\left.\begin{array}{ll}
\displaystyle
 z_1=\half\Big[\e^{-\i x}+(1+\e^{2y}-\e^{2z})\e^{\i x}\Big]
&\displaystyle
z_2=\frac{\i}{\sqrt{2}}\Big[\sinh(y-z)+\e^{y+z}\Big]\e^{\i x}
\\[3mm]
\displaystyle
 z_3=\frac{1}{\sqrt{2}}\Big[\sinh(y-z)-\e^{y+z}\Big]\e^{\i x}
&\displaystyle
z_4=\frac{\i}{2}\Big[\e^{-\i x}+(-1+\e^{2y}-\e^{2z})\e^{\i x}\Big]\e^{\i x}
\end{array}\quad\right\}
\end{equation}
$(x,y,z\in\bbbr)$. The set of commuting operators is given by
\begin{equation}
\CL_1=(I_{42}+\i I_{21})^2+(I_{34}+\i I_{13})^2,\qquad
\CL_2=I_{23}^2-(I_{42}+I_{31}+\i I_{12}+\i I_{34})^2\enspace.
\end{equation}
The metric terms are
\begin{equation}
\left.\begin{array}{l}
\d s^2=\d x^2+\e^{2\i x}(\e^{2y}+\e^{2z})(\d y^2-\d z^2)\enspace,
\\[3mm]
\sqrt{g}=\e^{2\i x}(\e^{2y}+\e^{2z})\enspace,
\\[3mm]
\displaystyle
\Gamma_x=2\i,\quad
\Gamma_y=\frac{2\e^{2y}}{\e^{2y}+\e^{2z}},\quad
\Gamma_z=\frac{-2\e^{2z}}{\e^{2y}+\e^{2z}}\enspace.
\end{array}\quad\right\}
\end{equation}
The momentum operators have the form
\begin{equation}
p_x=\hi\bigg(\Partial{x}+\i\bigg),\quad
p_{y}=\hi\bigg(\Partial{y}+\bhalf\Gamma_{y}\bigg),\quad
p_{z}=\hi\bigg(\Partial{z}+\bhalf\Gamma_{z}\bigg)\enspace,
\end{equation}
with the Hamiltonian given by
\begin{eqnarray} 
  H&=&-\hbarm\Bigg[\Partialsq{x}+2\i\Partial{\chi}
   +\frac{\e^{-2\i x}}{\e^{2y}+\e^{2z}}
   \Bigg(\Partialsq{y}+\Gamma_{y}\Partial{y}
  -\Partialsq{z}+\Gamma_{z}\Partial{z}\Bigg)\Bigg]\qquad
\nonumber\\   &=&
  {1\over2m}\left[p_x^2+\frac{\e^{-2\i x}}{\sqrt{\e^{2y}+\e^{2z}}}
   (p_{y}^2-p_{z}^2)
   \frac{1}{\sqrt{\e^{2y}-\e^{2z}}}\right]-\hbarm\enspace.
\end{eqnarray} 
For the path integral we obtain
\begin{eqnarray}
&&K^{(\SdreiC)}(x'',x',y'',y',z'',z';T)
\nonumber\\  &&
=\pathint{x}\pathint{y}\pathint{z}
\e^{2\i x}(\e^{2y}+\e^{2z})\qquad\qquad\qquad\qquad\qquad\qquad
\nonumber\\  &&\qquad\times
\exp\left\{\ih\int_0^T\Bigg[\frac{m}{2}\dot x^2
+\e^{2\i x}(\e^{2y}+\e^{2z})(\dot y^2-\dot z^2)+\frac{\hbar^2}{2m}
\Bigg]\dt\right\}\enspace.
\end{eqnarray}
We start by considering the $(y,z)$-subpath integration. This
two-dimensional sub-system corresponds to the second of the hyperbolic systems
on the two-dimensional pseudo-Euclidean plane $E(1,1)$ \cite{GROad,KALc,KAMIo},
in particular \cite{GROab}. 
The result is given by
\begin{eqnarray}
&&\hat K^{(E(1,1))}(y'',y',z'',z';T)
  ={2\over\pi^4}\int_0^\infty dk\,k\sinh\pi k\int_0^\infty dp\,p
         \nonumber\\   & &\qquad\qquad\qquad\qquad\times
   K_{\i k}(\e^{y_1''}p)K_{\i k}(\e^{y_1'}p)
   K_{\i k}(-\i\e^{y_2'}p)K_{\i k}(\i\e^{y_2''}p)
   \,\e^{-\i\hbar p^2T/2m}.\qquad\qquad
\end{eqnarray}
Using the result of the $x$-path integration of the horicyclic-polar
system we get finally
\begin{eqnarray}
&&K^{(\SdreiC)}(x'',x',y'',y',z'',z';T)
         \nonumber\\   & &
  ={2\over\pi^4}\int_0^\infty dk\,k\sinh\pi k\int_0^\infty dp\,p
   K_{\i k}(\e^{y''}p)K_{\i k}(\e^{y'}p)
   K_{\i k}(-\i\e^{z'}p)K_{\i k}(\i\e^{z''}p)
\nonumber\\  &&\qquad\times
\sum_{J\in\bbbn_0} \bhalf
H^{(1)}_{J+1/2}\big(px\e^{-\i x''}\big)
H^{(1)}_{J+1/2}\big(p\e^{\i x'}\big)
\exp\Bigg[-\ih\frac{\hbar^2J(J+2)}{2m}T\Bigg]\enspace.
\end{eqnarray}
The wave-functions on $\SdreiC$  are given by
\begin{equation}
\Psi_{Jpk}(x,y,z)=
\frac{ \sqrt{pk\sinh\pi k}}{\pi^2}
K_{\i k}(p\e^y)K_{\i k}(-\i p\e^z)
\e^{-\i x} H^{(1)}_{J+1/2}(p\,\e^{-\i x})\enspace,
\end{equation}
and the energy-spectrum (\ref{Energy-J}). 

\subsection{System 11: Horicyclic-Parabolic I}
\message{System 11: Horicyclic-Parabolic I}
This coordinate system is defined as
\begin{equation}
\left.\begin{array}{ll}
 z_1=\half\Big[\e^{-\i x}+(1+\bviert(\xi^2+\eta^2)^2)\e^{\i x}\Big]
&z_2=\frac{\i}{2}(\xi^2-\eta^2)\e^{\i x}
\\[3mm]
 z_3=\i\xi\eta\e^{\i x}               
&z_4=\frac{\i}{2}\Big[\e^{-\i x}+(-1+\bviert(\xi^2+\eta^2)^2)\e^{\i x}\Big]
\end{array}\quad\right\}
\end{equation}
$(x,\xi\in\bbbr,\eta>0)$, with the set of commuting operators given by
\begin{equation}
\CL_1=(I_{42}+\i I_{21})^2+(I_{34}+\i I_{13})^2,\qquad 
\CL_2=\{I_{23},I_{42}+\i I_{21}\}\enspace.
\end{equation}
The metric terms are
\begin{equation}
\left.\begin{array}{l}
\d s^2=\d x^2+\e^{2\i x}(\xi^2+\eta^2)(\d\xi^2+\d\eta^2)\enspace,
\\[3mm]
\sqrt{g}=\e^{2\i x}(\xi^2+\eta^2)\enspace,
\\[3mm]
\Gamma_x=2\i,\quad
\Gamma_\xi=\dfrac{2\xi}{\xi^2+\eta^2},\quad
\Gamma_\eta=\dfrac{2\eta}{\xi^2+\eta^2}\enspace.
\end{array}\quad\right\}
\end{equation}
The momentum operators read
\begin{equation}
p_x=\hi\bigg(\Partial{x}+\i\bigg),\quad
p_{\xi}=\hi\bigg(\Partial{\xi}+\frac{\xi}{\xi^2+\eta^2}\bigg),\quad
p_{\eta}=\hi\bigg(\Partial{\eta}+\frac{\eta}{\xi^2+\eta^2}\bigg)\enspace,
\end{equation}
and the Hamiltonian has the form
\begin{eqnarray} 
  H&=&-\hbarm\Bigg[\Partialsq{x}+2\i\Partial{\chi}
   +\frac{\e^{-2\i x}}{\xi^2+\eta^2}
   \Bigg(\Partialsq{\xi}+\frac{2\xi}{\xi^2+\eta^2}\Partial{\xi}
  +\Partialsq{z}+\frac{2\eta}{\xi^2+\eta^2}\Partial{\eta}\Bigg)\Bigg]\qquad
\nonumber\\   &=&
  {1\over2m}\left[p_x^2+\frac{\e^{-2\i x}}{\sqrt{\xi^2+\eta^2}}
   (p_{\xi}^2+p_{\eta}^2)
   \frac{1}{\sqrt{\xi^2+\eta^2}}\right]-\hbarm\enspace.
\end{eqnarray} 
For the path integral representation we obtain
\begin{eqnarray}
&&K(x'',x',\xi'',\xi',\eta'',\eta';T)
\nonumber\\  &&
=\pathint{x}\pathint{\xi}\pathint{\eta}\e^{2\i x}(\xi^2+\eta^2)
\nonumber\\  &&\qquad\times
\exp\left\{\ih\int_0^T\left[\frac{m}{2}
\Big(\d x^2+\e^{2\i x}(\xi^2+\eta^2)(\d\xi^2+\d\eta^2)\Big)
+\frac{\hbar^2}{2m}\right]\dt\right\}
\nonumber\\  &&
=\int_{\bbbr} d\zeta\int_{\bbbr}{dp\over32\pi^4}
         \nonumber\\   & &\qquad\times
 \left\{\begin{array}{l}\big|\Gamma(\viert+{\i\zeta\over2p})\big|^2
    E^{(0)}_{-1/2+\i\zeta/p}(\e^{-\i\pi/4}\sqrt{2p}\,\xi'')
    E^{(0)}_{-1/2-\i\zeta/p}(\e^{-\i\pi/4}\sqrt{2p}\,\eta'')    \\
    \big|\Gamma({3\over4}+{\i\zeta\over2p})\big|^2
    E^{(1)}_{-1/2+\i\zeta/p}(\e^{ \i\pi/4}\sqrt{2p}\,\xi'')
    E^{(1)}_{-1/2-\i\zeta/p}(\e^{ \i\pi/4}\sqrt{2p}\,\eta'')
    \end{array}\right\}\qquad
         \nonumber\\   & &\qquad\times
    \left\{\begin{array}{l}\big|\Gamma(\viert+{\i\zeta\over2p})\big|^2
    E^{(0)}_{-1/2-\i\zeta/p}(\e^{ \i\pi/4}\sqrt{2p}\,\xi')
    E^{(0)}_{-1/2+\i\zeta/p}(\e^{ \i\pi/4}\sqrt{2p}\,\eta')     \\
    \big|\Gamma({3\over4}+{\i\zeta\over2p})\big|^2
    E^{(1)}_{-1/2-\i\zeta/p}(\e^{-\i\pi/4}\sqrt{2p}\,\xi')
    E^{(1)}_{-1/2+\i\zeta/p}(\e^{-\i\pi/4}\sqrt{2p}\,\eta')
    \end{array}\right\}
\nonumber\\  &&\qquad\times
\sum_{J\in\bbbn_0} \bhalf
H^{(1)}_{J+1/2}\big(p\e^{-\i x''}\big)
H^{(1)}_{J+1/2}\big(p\e^{\i x'}\big)
\exp\Bigg[-\ih\frac{\hbar^2J(J+2)}{2m}T\Bigg]\enspace.
\end{eqnarray}
In this solution we have exploited the path integral representation on
the two-dimensional Euclidean plane in parabolic coordinates
\cite{GROab,GROad}, in particular \cite{GROab} for details.
The wave-functions on $\SdreiC$  $E^{(0)}_{-1/2-\i\zeta/p}(z)$ are of even
parity, whereas 
the wave-functions $E^{(1)}_{-1/2-\i\zeta/p}(z)$ are of odd parity.
The wave-functions on $\SdreiC$  for even and odd parity, respectively, are given by
\begin{eqnarray}
&&\Psi_{Jp\zeta}(x,\xi,\eta)=
{1\over8\pi^2}
         \nonumber\\   & &\qquad\times
    \left\{\begin{array}{l}
    \big|\Gamma(\viert+{\i\zeta\over2p})\big|^2
    E^{(0)}_{-1/2+\i\zeta/p}(\e^{-\i\pi/4}\sqrt{2p}\,\xi)
    E^{(0)}_{-1/2-\i\zeta/p}(\e^{-\i\pi/4}\sqrt{2p}\,\eta)    \\
    \big|\Gamma({3\over4}+{\i\zeta\over2p})\big|^2
    E^{(1)}_{-1/2+\i\zeta/p}(\e^{ \i\pi/4}\sqrt{2p}\,\xi)
    E^{(1)}_{-1/2-\i\zeta/p}(\e^{ \i\pi/4}\sqrt{2p}\,\eta)
    \end{array}\right\}\qquad
         \nonumber\\   & &\qquad\times
\e^{-\i x}\,H^{(1)}_{J+1/2}(p\,\e^{-i x})\enspace,
\end{eqnarray}
and the energy-spectrum (\ref{Energy-J}). 

\subsection{System 12: Horicyclic-Parabolic II}
\message{System 12: Horicyclic-Parabolic II}
This coordinate system is defined as
\begin{equation}
\left.\begin{array}{ll}
 z_1=\half\Big\{\e^{-\i x}+[1+2(\xi-\eta)^2(\xi+\eta)]\e^{\i x}\Big\}
&z_2=\i\big[\bhalf(\xi-\eta)^2+(\xi+\eta)\big]\e^{\i x}
\\[3mm]
 z_3=\big[\bhalf(\xi-\eta)^2-(\xi+\eta)\big]\e^{\i x}           
&z_4=\half\Big\{\e^{-\i x}+[-1+2(\xi-\eta)^2(\xi+\eta)]\e^{\i x}\Big\}.
\end{array}\quad\right\}
\end{equation}
The set of commuting operators is given by
\begin{equation}
\CL_1= (I_{12}+\i I_{21})^2+(I_{34}+\i I_{13})^2,\qquad 
\CL_2=\{I_{23},I_{42}+I_{31}+\i I_{21}+\i I_{34}\}
    -\i(I_{42}-I_{31}+\i I_{21}-\i I_{34})^2\enspace.
\end{equation}
The metric terms are
\begin{equation}
\left.\begin{array}{l}
\d s^2=\d x^2+4\e^{2\i x}(\xi-\eta)(\d\xi^2-\d\eta^2)\enspace,
\\[3mm]
\sqrt{g}=4\e^{2\i x}(\xi-\eta)\enspace,
\\[3mm]
\Gamma_x=2\i,\quad
\Gamma_\xi=\dfrac{1}{\xi-\eta},\quad
\Gamma_\eta=\dfrac{-1}{\xi-\eta}\enspace.
\end{array}\quad\right\}
\end{equation}
The momentum operators have the form
\begin{equation}
p_x=\hi\bigg(\Partial{x}+\i\bigg),\quad
p_{\xi}=\hi\bigg(\Partial{\xi}+\frac{1}{2(\xi-\eta)}\bigg),\quad
p_{\eta}=\hi\bigg(\Partial{\eta}-\frac{1}{2(\xi-\eta)}\bigg)\enspace,
\end{equation}
and the Hamiltonian reads
\begin{eqnarray} 
  H&=&-\hbarm\Bigg[\Partialsq{x}+2\i\Partial{\chi}
   +\frac{\e^{-2\i x}}{\xi-\eta}
   \Bigg(\Partialsq{\xi} +\frac{1}{\xi-\eta}\Partial{\xi}
        -\Partialsq{\eta}+\frac{1}{\xi-\eta}\Partial{\eta}\Bigg)\Bigg]\qquad
\nonumber\\   &=&
  {1\over2m}\left[p_x^2+\frac{\e^{-2\i x}}{\sqrt{\xi-\eta}}
   (p_{\xi}^2-p_{\eta}^2)
   \frac{1}{\sqrt{\xi-\eta}}\right]-\hbarm\enspace.
\end{eqnarray} 
For the path integral we find
\begin{eqnarray}
&&K^{(\SdreiC)}(x'',x',\xi'',\xi',\eta'',\eta';T)
\nonumber\\  &&
=\pathint{x}\pathint{\xi}\pathint{\eta}4\e^{2\i x}(\xi-\eta)
\nonumber\\  &&\qquad\times
\exp\left\{\ih\int_0^T\Bigg[\frac{m}{2}
\dot x^2+4\e^{2\i  x}(\xi-\eta)(\dot\xi^2-\dot\eta^2)+\frac{\hbar^2}{2m}
\Bigg]\dt\right\}\enspace.\qquad\qquad
\end{eqnarray}
The $(\xi,\eta)$-subpath integration corresponds to the path integration
of the third parabolic system on the two-dimensional pseudo-Euclidean
plane \cite{GROad,KALc,KAMIo}. Using the result of \cite{GROad} and
the horicyclic system we get
\begin{eqnarray}
&&K^{(\SdreiC)}(x'',x',\xi'',\xi',\eta'',\eta';T)
\nonumber\\  &&
  =16\int_0^\infty{dp\over p^{1/3}}
   \int_{\bbbr}d\zeta\,
  \Ai\Bigg[-\bigg(\xi'+\sqrt{2m}{\zeta\over p^2}\bigg)p^{2/3}\Bigg]
  \Ai\Bigg[-\bigg(\xi''+\sqrt{2m}{\zeta\over p^2}\bigg)p^{2/3}\Bigg]
         \nonumber\\   & &\qquad\times
  \Ai\Bigg[-\bigg(\eta'+\sqrt{2m}{\zeta\over p^2}\bigg)p^{2/3}\Bigg]
  \Ai\Bigg[-\bigg(\eta''+\sqrt{2m}{\zeta\over p^2}\bigg)p^{2/3}\Bigg]
\nonumber\\  &&\qquad\times
\sum_{J\in\bbbn_0} \bhalf
H^{(1)}_{J+1/2}\big(p\e^{-\i x''}\big)
H^{(1)}_{J+1/2}\big(px\e^{\i x'}\big)
\exp\Bigg[-\ih\frac{\hbar^2J(J+2)}{2m}T\Bigg]\enspace.
\end{eqnarray}
Here, the $\Ai$ are Airy-functions \cite{GRA}.
The wave-functions on $\SdreiC$  are given by
\begin{eqnarray}
&&\Psi_{Jp\zeta}(x,\xi,\eta)=
{4\over p^{1/6}}
\Ai\Bigg[-\bigg(\xi+\sqrt{2m}{\zeta\over p^2}\bigg)p^{2/3}\Bigg]
\Ai\Bigg[-\bigg(\eta+\sqrt{2m}{\zeta\over p^2}\bigg)p^{2/3}\Bigg]
\nonumber\\  &&\qquad\times
\frac{1}{\sqrt{2}}\,\e^{-\i x}H^{(1)}_{J+1/2}(p\,\e^{-\i x})\enspace,
\end{eqnarray}
and the energy-spectrum (\ref{Energy-J}). 

\setcounter{equation}{0}
\section{The Path Integral Representations: Part II}
\message{The Path Integral Representations: Part II}

\subsection{System 13: Elliptic-Cylindrical}
\message{System 13: Elliptic-Cylindrical}
We now come the those coordinate systems which do not have a subgroup
structure. There are nine of them, and we can find for six of these cases a
path integral representation.
 
This coordinate system of the elliptic-cylindrical type is defined as
\begin{equation}
\left.\begin{array}{ll}
\displaystyle
 z_1=k\sn(\alpha,k)\sn(\beta,k)
&\displaystyle
 z_2=-\i\frac{k}{k'}\cn(\alpha,k)\cn(\beta,k)\cos\vphi
\\[3mm]
\displaystyle
 z_3=-\i\frac{k}{k'}\cn(\alpha,k)\cn(\beta,k)\sin\vphi                 
&\displaystyle
z_4=\frac{1}{k'}\dn(\alpha,k)\dn(\beta,k)
\end{array}\quad\right\}\enspace,
\end{equation}
The set of commuting operators is given by
\begin{equation}
\CL_1=I_{23}^2,\qquad 
\CL_2=I_{12}^2+I_{13}^2+kI_{14}^2\enspace.
\end{equation}
and the metric terms are
\begin{equation}
\left.\begin{array}{l}
\d s^2=-k^2(\sn^2\alpha-\sn^2\beta)(\d\alpha^2+\d\beta^2)
+\dfrac{k^2}{{k'}^2}\cn^2\alpha\cn^2\beta\d\vphi^2\enspace,
\\
\sqrt{g}=k^2(\sn^2\alpha-\sn^2\beta)\dfrac{k}{{k'}}\,\cn\alpha\cn\beta\enspace,
\\
\Gamma_\alpha=-\dfrac{2k^2\sn\alpha\cn\alpha\dn\alpha}{\emetric}
         +\dfrac{\cn\alpha\dn\alpha}{\sn\alpha},\quad
\Gamma_\beta=-\dfrac{2{k'}^2\sn\beta\cn\beta\dn\beta}{\emetric}
         -{k'}^2\dfrac{\sn\beta\cn\beta}{\dn\beta}\enspace,
\end{array}\quad\right\}
\end{equation}
and $\Gamma_\vphi=0$.
In \cite{GROad,GROPOb} we have constructed a kernel for the prolate
elliptic coordinate system on the sphere $S^{(3)}$. We used the definition
for the coordinates ($a_1\leq\rho_1\leq a_2\leq\rho_2\leq a_3$,
algebraic from)
\begin{equation}\left.\begin{array}{l}
  s_1^2=R^2\dfrac{(\rho_1-a_2)(\rho_2-a_2)}{(a_3-a_2)(a_1-a_2)}
  \cos^2\vphi\enspace,         \\ \vphantom{\Bigg)}
  s_2^2=R^2\dfrac{(\rho_1-a_2)(\rho_2-a_2)}{(a_3-a_2)(a_1-a_2)}
  \sin^2\vphi\enspace,         \\ \vphantom{\Bigg)}
  s_3^2=R^2\dfrac{(\rho_1-a_1)(\rho_2-a_1)}{(a_2-a_1)(a_3-a_1)}
  \enspace,                   \\
  s_4^2=R^2\dfrac{(\rho_1-a_3)(\rho_2-a_3)}{(a_2-a_3)(a_1-a_3)}
  \enspace.                   \end{array}\qquad\qquad\right\}
\end{equation}
In terms of the Jacobi elliptic functions we have
($-K\leq \alpha\leq K, -2K'\leq\beta\leq2K', 0\leq\vphi<2\pi$)
\begin{equation}
\left.\begin{array}{l}
  s_1=R\cn(\alpha,k)\cn(\beta,k')\cos\vphi\enspace,   \\
  s_2=R\cn(\alpha,k)\cn(\beta,k')\sin\vphi\enspace,   \\
  s_3=R\sn(\alpha,k)\dn(\beta,k')\enspace,    \\
  s_4=R\dn(\alpha,k)\sn(\beta,k')\enspace.
  \end{array}\qquad\qquad\qquad\right\}
\end{equation}
The momentum operators are given by
\begin{equation}
  p_\alpha=\hi\bigg({\partial\over\partial\alpha}
         +\bhalf\Gamma_\alpha\bigg),\quad
  p_\beta=\hi\bigg({\partial\over\partial\beta}
         +\bhalf\Gamma_\alpha\bigg),\quad
  p_\vphi=\hi{\partial\over\partial\vphi}       \enspace.
\end{equation}
Therefore we have for the Hamiltonian
\begin{eqnarray}
   H
   &=&-\hbarm\Bigg[
   {1\over\Emetric}\bigg({\partial^2\over\partial\alpha^2}
       +{\cn\alpha\dn\alpha\over\sn\alpha}{\partial\over\partial\alpha}
         \nonumber\\   & &
   \qquad\qquad\qquad\qquad\qquad
        +{\partial^2\over\partial\beta^2}
       -{k'}^2{\sn\beta\cn\beta\over\dn\beta}{\partial\over\partial\beta}
   \bigg)+{1\over\sn^2\alpha\dn^2\beta}{\partial^2\over\partial\vphi^2}\Bigg]
                  \\
  &=&\hbarm{1\over\sqrt{\emetric}}
   (p_\alpha^2+p_\beta^2){1\over\sqrt{\emetric}}
         \nonumber\\   & &\qquad
 -{\hbar^2\over8m}\Bigg[4+{1\over\emetric}
   \bigg({\cn^2\alpha\dn^2\alpha\over\sn^2\alpha}
   +{k'}^4{\sn^2\beta\cn^2\beta\over\dn^2\beta}\bigg)\Bigg]\enspace.\qquad
\end{eqnarray}
We found the following representation \cite{GROad} ($\alpha\in[-K,K],
\beta\in[-K',K'], \vphi\in[0,2\pi),a\in[-1,0]$):
\begin{eqnarray}
&&K^{(\SdreiC)}(\alpha'',\alpha',\beta'',\beta',\vphi'',\vphi';T)
\nonumber\\  &&
=\pathint\alpha\pathint\beta\Emetric\cn\alpha\cn\beta\pathint\vphi
         \nonumber\\   & &\qquad\times
   \exp\Bigg\{\ih\intt\Bigg[{m\over2}\Big(\Emetric
   (\dot\alpha^2+\dot\beta^2)+\cn^2\alpha\cn^2\beta\dot\vphi^2\Big)
         \nonumber\\   & &
   \qquad\qquad\qquad\qquad
    +\hbaram{1\over\emetric}\bigg({\sn^2\beta\dn^2\beta\over\cn^2\beta}
    +k^4{\sn^2\alpha\dn^2\alpha\over\cn^2\alpha}\bigg)
    \Bigg]dt+{\i\hbar T\over2m}\Bigg\}
         \nonumber\\   & &
   ={1\over2\pi}\sum_{J=0}^\infty\sum_{r,p=\pm1}\sum_{qk_2}
   \e^{\i k_2(\vphi''-\vphi')}\e^{-2\i\hbar TJ(J+2)/2m}
         \nonumber\\   & &\qquad\times
  \psi_{1,Jqk_2}^{(r,p)}(\alpha'';a)\psi_{1,Jqk_2}^{(r,p)\,*}(\alpha';a)
  \psi_{2,Jqk_2}^{(r,p)}(\beta'';a)\psi_{2,Jqk_2}^{(r,p)\,*}(\beta';a)
    \vphantom{\bigg]}\enspace.
\label{numEe}
\end{eqnarray}
This representation can be derived by a group path
integration together with an interbasis expansion from the cylindrical
or spherical basis to the ellipso-cylindrical ones. For instance, one
has \cite{GKPSa} $(r,p=\pm1)$
\begin{eqnarray}
 \Psi_{J,k_1,k_2}^{(r,p)}(\vtheta,\vphi_1,\vphi_2)&=&\sum_q
 T_{Jqk_2}^{(r,p)\,*}\Psi_{Jqk_2}^{(r,p)}(\alpha,\beta,\vphi;a)\enspace,
\label{numEg}
\end{eqnarray}
where the $a\geq0$ corresponds to the oblate elliptic case (which we do not
discuss here), and $a\in[-1,0]$
to the prolate elliptic system, respectively. We have the factorization 
$\Psi(\alpha,\beta,
\vphi;a)=\psi_1(\alpha;a)\psi_2(\beta;a)\e^{\i k_2\vphi}/\sqrt{2\pi}$.
For the associated Lam\'e polynomials $\psi_{i,Jqk_2}$, $i=1,2$, we adopt
the notations of \cite{GKPSa,KMW}. The relevant quantum numbers have the
following meaning: The functions $\psi_{i,Jqk_2}^{(r,p)}(z)$ are
called {\it associated Lam\'e polynomials} and satisfy the associated Lam\'e
equation. We take them for normalized. For the principal quantum number
we have $l\in\bbbn_0$. $(r,p)=\pm1$ denotes one of the four parity
classes of solutions of dimension $(J+1)^2$, i.e., the multiplicity of
the degeneracy of the level $J$, one for each class of the corresponding
recurrence relations as given in \cite{GKPSa,KMW} and the parity classes
from the periodic Lam\'e functions $\Lambda_{lh}^p$ from the spherical
harmonics on the sphere can be applied. These expansions have been
considered in \cite{GKPSa,KMW} together with three-term recurrence
relations for the interbasis coefficients. They can be determined by
taking into account that a basis in $\OO(4)$ is related in a unique way
to the cylindrical and spherical bases on $\sdrei$ by using the
properties of the elliptic operator $\Lambda$ on $\sdrei$ with
eigenvalue $q$. Details can be found in \cite{GKPSa,KMW}. Due to the
unitarity of these coefficients the path integration is then performed
by inserting in each short-time kernel in the cylindrical system first
the expansions (\ref{numEg}), second, exploiting the unitarity, and thus
yielding the result (\ref{numEe}).
The elliptic-cylindrical system exists on the three-dimensional sphere
(oblate and prolate spheroidal (System IV and V.), 
on the three-dimensional hyperboloid 
(System XVII. and XVIII -- prolate and oblate elliptic).

\subsection{System 14: Elliptic-Parabolic}
\message{System 14: Elliptic-Parabolic}
This coordinate system is defined as
\begin{equation}
\left.\begin{array}{ll}
\displaystyle
 z_1=\half\bigg(\frac{\cosh\tau_1}{\cosh\tau_2}
               +\frac{\cosh\tau_2}{\cosh\tau_1}\bigg)
&z_2=\tanh\tau_1\tanh\tau_2\cosh\tau_3
\\[3mm]
\displaystyle
 z_3=-\i\tanh\tau_1\tanh\tau_2\sinh\tau_3
&\displaystyle
z_4=\frac{-\i}{\cosh\tau_1\cosh\tau_2}+
\frac{\i}{2}\bigg(\frac{\cosh\tau_1}{\cosh\tau_2}
               +\frac{\cosh\tau_2}{\cosh\tau_1}\bigg)
\end{array}\quad\right\}\enspace,
\end{equation}
$(\tau_1,\tau_2>0,\tau_3\in\bbbr)$.
The set of commuting operators is given by
\begin{equation}
\CL_1=I_{23}^2,\qquad 
\CL_2=I_{24}^2+I_{34}^2-I_{12}^2+I_{13}^2-I_{14}^2
-\i\{I_{12},I_{42}\}-\i\{I_{13},I_{43}\}\enspace.
\end{equation}
The metric terms are
\begin{equation}
\left.\begin{array}{l}
\d s^2=(\tanh^2\tau_1-\tanh^2\tau_2)(\d\tau_1^2-\d\tau_2^2)
+\tanh^2\tau_1\tanh^2\tau_2\d\tau_3^2\enspace,
\\[3mm]
\sqrt{g}=(\tanh^2\tau_1-\tanh^2\tau_2)\tanh\tau_1\tanh\tau_2\enspace,
\\[3mm]
\Gamma_{\tau_{1,2}}=\pm\dfrac{2\sinh\tau_{1,2}}{\cosh^3\tau_{1,2}}
\dfrac{1}{\tanh^2\tau_1-\tanh^2\tau_2}+\dfrac{1}{\cosh^2\tau_{1,2}},\quad
\Gamma_{\tau_3}=0\enspace.
\end{array}\quad\right\}
\end{equation}
The momentum operators have the form
\begin{equation}
p_{\tau_1}=\hi\bigg(\Partial{\tau_1}+\bhalf\Gamma_{\tau_1}\bigg),\quad
p_{\tau_2}=\hi\bigg(\Partial{\tau_2}+\bhalf\Gamma_{\tau_2}\bigg),\quad
p_{\tau_3}=\hi\Partial{\tau_3}\enspace,
\end{equation}
with the Hamiltonian
\begin{eqnarray} 
  H&=&-\hbarm\Bigg[\frac{1}{\tanh^2\tau_1-\tanh^2\tau_2}\Bigg(
  \Partialsq{\tau_1}+\Gamma_{\tau_1}\Partial{\tau_1}
  -\Partialsq{\tau_2}+\Gamma_{\tau_2}\Partial{\tau_2}\Bigg)
   +\frac{1}{\tanh^2\tau_1\tanh^2\tau_2}\Partialsq{\tau_3}\Bigg]
\nonumber\\   &=&
  {1\over2m}\left[\frac{1}{\sqrt{\tanh^2\tau_1-\tanh^2\tau_2}}
   (p_{\tau_1}^2-p_{\tau_2}^2)
   \frac{1}{\sqrt{\tanh^2\tau_1-\tanh^2\tau_2}}
  +\frac{p_{\tau_3}^2}{\tanh^2\tau_1\tanh^2\tau_2}\right]
  +\frac{3\hbar^2}{8m}\,.
\nonumber\\   &&
\end{eqnarray} 
The solution cannot be directly derived from one of the 
lower-dimensional subcases. For the path integral we find 
by separating off the $\tau_3$-path integration which corresponds to plane waves
\begin{eqnarray}
&&K^{(\SdreiC)}(\tau_1'',\tau_1',\tau_2'',\tau_2',\tau_3'',\tau_3';T)
\nonumber\\  &&
=\pathint{\tau_1}\pathint{\tau_2}\pathint{\tau_3}
(\tanh^2\tau_1-\tanh^2\tau_2)\tanh\tau_1\tanh\tau_2
\nonumber\\  &&\qquad\times
\exp\left\{\ih\int_0^T\Bigg[\frac{m}{2}\left(
(\tanh^2\tau_1-\tanh^2\tau_2)(\dot\tau_1^2-\dot\tau_2^2)
+\tanh^2\tau_1\tanh^2\tau_2\dot\tau_3^2\right)-\frac{3\hbar^2}{8m}
\Bigg]\dt\right\}
\nonumber\\  &&
=(\tanh\tau_1'\tanh\tau_1''\tanh\tau_2'\tanh\tau_2'')^{-1/2}
\int_{\bbbr}\d k_{\tau_3}\frac{\e^{\i k_{\tau_3}(\tau_3''-\tau_3')}}{2\pi}
\nonumber\\  &&\qquad\times
\pathint{\tau_1}\pathint{\tau_2}(\tanh^2\tau_1-\tanh^2\tau_2)
\nonumber\\  &&\qquad\times
\exp\left\{\ih\int_0^T\Bigg[\frac{m}{2}
(\tanh^2\tau_1-\tanh^2\tau_2)(\dot\tau_1^2-\dot\tau_2^2)
-\frac{\hbar^2k_{\tau_3}^2}{2m}\coth^2\tau_1\coth^2\tau_2-\frac{3\hbar^2}{8m}
\Bigg]\dt\right\}
\nonumber\\  &&
=(\tanh\tau_1'\tanh\tau_1''\tanh\tau_2'\tanh\tau_2'')^{-1/2}
\nonumber\\  &&\qquad\times
\int_{\bbbr}\d k_{\tau_3}\frac{\e^{\i k_{\tau_3}(\tau_3''-\tau_3')}}{2\pi}
\int_{\bbbr}\frac{\d E}{2\pi\hbar}\e^{\i ET/\hbar}\int_0^\infty \d s''
K_k(\tau_1'',\tau_1',\tau_2'',\tau_2';s'')\enspace,
\end{eqnarray}
with the space-time transformed path integral $K_k(s'')$ given by:
\begin{eqnarray}
&&K_k(\tau_1'',\tau_1',\tau_2'',\tau_2';s'')
=\pathints{\tau_1}\pathints{\tau_2}
\nonumber\\  &&\qquad\times
\exp\left\{\ih\int_0^{s''}\Bigg[\frac{m}{2}(\dot\tau_1^2-\dot\tau_2^2)
+\bigg(E+\frac{3\hbar^2}{8m}\bigg)
\bigg(\frac{1}{\cosh^2\tau_1}-\frac{1}{\cosh^2\tau_2}\bigg)
\right.
\nonumber\\  &&\qquad\qquad\qquad\qquad\qquad\qquad\qquad\qquad\left.
+\frac{\hbar^2k_{\tau_3}^2}{2m}
\bigg(\frac{1}{\sinh^2\tau_1}-\frac{1}{\sinh^2\tau_2}\bigg)
\Bigg]\d s\right\}. \qquad 
\end{eqnarray}
For this path integral we now use the path integral representation of the
modified P\"oschl--Teller potential
\begin{equation}
V^{(mPT)}(\tau)=\frac{\hbar^2}{2m}\left(\frac{\eta^2-\viert}{\sinh^2\tau}
                -\frac{\nu^2-\viert}{\cosh^2\tau}\right)\enspace.
\end{equation}
We therefore obtain with $\eta=\i\sqrt{k_{\tau_3}^2-\viert}$ and
$\nu=2mE/\hbar^2+1\equiv p^2+1$
\begin{eqnarray}
&&\!\!\!\!\!\!\!
  K^{(\SdreiC)}(\tau_1'',\tau_1',\tau_2'',\tau_2',\tau_3'',\tau_3';T)
=(\tanh\tau_1'\tanh\tau_1''\tanh\tau_2'\tanh\tau_2'')^{-1/2}
\nonumber\\  &&\!\!\!\!\!\!\!\qquad\qquad\times
\int_{\bbbr}\d k\frac{\e^{\i k_{\tau_3}(\tau_3''-\tau_3')}}{2\pi}
\int_0^\infty dk\int_0^\infty dp\,\Energyldrei\qquad\qquad
         \nonumber\\   & &\!\!\!\!\!\!\!\qquad\qquad\times
    \Psi_k^{(\eta,\nu)}(\tau_2'')\Psi_k^{(\eta,\nu)\,*}(\tau_2')
    \Psi_k^{(\eta,\nu)}(\tau_1'')\Psi_k^{(\eta,\nu)\,*}(\tau_1')\enspace.
\end{eqnarray}
In this solution we have explicitly $J(J+2)\equiv(p^2+1)$.
The $\Psi_p^{(\mu,\nu)}(\omega)$ are the modified  P\"oschl--Teller
functions, which are given by
\begin{eqnarray}
  \Psi_n^{(\eta,\nu)}(r)
  &=&N_n^{(\eta,\nu)}(\sinh r)^{2k_2-\half}
                    (\cosh r)^{-2k_1+{3\over2 }}
  \nonumber\\   &&\qquad\times
  {_2}F_1(-k_1+k_2+\kappa,-k_1+k_2-\kappa+1;2k_2;-\sinh^2r)
  \\
  N_n^{(\eta,\nu)}
  &=&{1\over\Gamma(2k_2)}
  \bigg[{2(2\kappa-1)\Gamma(k_1+k_2-\kappa)
                     \Gamma(k_1+k_2+\kappa-1)\over
    \Gamma(k_1-k_2+\kappa)\Gamma(k_1-k_2-\kappa+1)}\bigg]^{1/2}\enspace.
\end{eqnarray}
The scattering states are given by:
 \begin{eqnarray}
  V(r)&=&\hbarm \bigg({\eta^2-{1\over4}\over\sinh^2r}
   -{\nu^2-{1\over4}\over\cosh^2r}\bigg)
  \nonumber\\   
  \Psi_p^{(\eta,\nu)}(r)
  &=&N_p^{(\eta,\nu)}(\cosh r)^{2k_1-\half}(\sinh r)^{2k_2-\half}
  \nonumber\\   &&\qquad\qquad\times
  {_2}F_1(k_1+k_2-\kappa,k_1+k_2+\kappa-1;2k_2;-\sinh^2r)
  \\
  N_p^{(\eta,\nu)}
  &=&{1\over\Gamma(2k_2)}\sqrt{p\sinh\pi p\over2\pi^2}
  \Big[\Gamma(k_1+k_2-\kappa)\Gamma(-k_1+k_2+\kappa)
  \nonumber\\   &&\qquad\qquad\times
  \Gamma(k_1+k_2+\kappa-1)\Gamma(-k_1+k_2-\kappa+1)\Big]^{1/2}\enspace,
\end{eqnarray}
$k_1,k_2$ defined by:
$k_1=\half(1\pm\nu)$, $k_2=\half(1\pm\eta)$, where the correct sign
depends on the boundary-conditions for $r\to0$ and $r\to\infty$,
respectively. The number $N_M$ denotes the maximal number of
states with $0,1,\dots,N_M<k_1-k_2-\half$. $\kappa=k_1-k_2-n$ for the
bound states and $\kappa=\half(1+\i p)$ for the scattering states.
${_2}F_1(a,b;c;z)$ is the hypergeometric function \cite[p.1057]{GRA}.
The bound states are needed for the bound states on the O(2,2) hyperboloid.

\subsection{System 15: Elliptic-Hyperbolic}
\message{System 15: Elliptic-Hyperbolic}
This coordinate system is defined as
\begin{equation}
\left.\begin{array}{l}
\displaystyle
z_1=-\half\bigg(\frac{\cosh\tau_2}{\cosh\tau_1}
                +\frac{\cosh\tau_1}{\cosh\tau_2}\bigg)
     -\frac{\tau_3^2}{2\cosh\tau_1\cosh\tau_2}\\[3mm]
\displaystyle
z_2=\frac{\i\tau_3}{\cosh\tau_1\cosh\tau_2}
\\[3mm]
\displaystyle
 z_3=\tanh\tau_1\tanh\tau_2\\[3mm]
\displaystyle
z_4=\i\bigg[\frac{2-\tau_3^2}{2\cosh\tau_1\cosh\tau_2}
           -\half\bigg(\frac{\cosh\tau_1}{\cosh\tau_2}
                +\frac{\cosh\tau_2}{\cosh\tau_1}\bigg)\bigg]
\end{array}\quad\right\}\enspace,
\end{equation}
($\tau_1,\tau_2,\tau_3\in\bbbr$). The set of commuting operators is given by
\begin{equation}
\CL_1=(I_{42}+\i I_{21})^2,\qquad 
\CL_2=2I_{12}^2+I_{13}^2++I_{14}^2-I_{34}^2
+\i(\{I_{12},I_{42}\}+\{I_{13},I_{43}\})\enspace.
\end{equation}
The metric terms are given by
\begin{equation}
\left.\begin{array}{l}
\displaystyle
\d s^2=\bigg(\frac{1}{\cosh^2\tau_1}-\frac{1}{\cosh^2\tau_2}\bigg)
       (\d\tau_1^2-\d\tau_2^2)+\frac{1}{\cosh^2\tau_1\cosh^2\tau_2}
       \d\tau_3^2\enspace,
\\[3mm]
\displaystyle
\sqrt{g}=\bigg(\frac{1}{\cosh^2\tau_1}-\frac{1}{\cosh^2\tau_2}\bigg)
\frac{1}{\cosh\tau_1\cosh\tau_2}\enspace,
\\[3mm]
\Gamma_{\tau_1}=\dfrac{-2\sinh\tau_1}{\cosh^3\tau_1}
\dfrac{1}{{1}/{\cosh^2\tau_1}-{1}/{\cosh^2\tau_2}}-\tanh\tau_1
\enspace,\\
\Gamma_{\tau_2}=\dfrac{2\sinh\tau_2}{\cosh^3\tau_2}
\dfrac{1}{{1}/{\cosh^2\tau_1}-{1}/{\cosh^2\tau_2}}-\tanh\tau_2
\enspace,
\end{array}\quad\right\}
\end{equation}
and $\Gamma_{\tau_3}=0$. We have for the momentum operators
\begin{equation}
p_{\tau_1}=\hi\left(\Partial{\tau_1}+\bhalf\Gamma_{\tau_1}\right),\qquad
p_{\tau_2}=\hi\left(\Partial{\tau_2}+\bhalf\Gamma_{\tau_2}\right),\qquad
p_{\tau_3}=\hi\Partial{\tau_3}\enspace.
\end{equation}
This gives for the Hamiltonian
\begin{eqnarray}
\!\!\!\!\!\!\!\!\!\!\!\!\!\!\!\!
&&H=-\frac{\hbar^2}{2m}\Bigg[
\frac{\cosh^2\tau_1\cosh^2\tau_2}{\cosh^2\tau_2-\cosh^2\tau_1}
\left(\Partialsq{\tau_1}-\tanh\tau_1\Partial{\tau_1}
-\Partialsq{\tau_2}+\tanh\tau_2\Partial{\tau_2}\right)
\nonumber\\  &&\qquad\qquad\qquad\qquad\qquad\qquad
\qquad\qquad\qquad\qquad\qquad\qquad
+\cosh^2\tau_1\cosh^2\tau_2\Partialsq{\tau_3}\Bigg]
\nonumber\\
\!\!\!\!\!\!\!\!\!\!\!\!\!\!\!\!
&&=\frac{1}{2m}\left[
\sqrt{\frac{\cosh^2\tau_1\cosh^2\tau_2}{\cosh^2\tau_2-\cosh^2\tau_1}}
\big(p_{\tau_1}^2-p_{\tau_2}^2\big)
\sqrt{\frac{\cosh^2\tau_1\cosh^2\tau_2}{\cosh^2\tau_2-\cosh^2\tau_1}}
+\cosh^2\tau_1\cosh^2\tau_2p_{\tau_3}^2\right]
+\frac{3\hbar^2}{8m}. 
\nonumber\\
\end{eqnarray}
For the path integral we find 
\begin{eqnarray}
&&\!\!\!\!\!\!\!\!
K^{(\SdreiC)}(\tau_1'',\tau_1',\tau_2'',\tau_2',\tau_3'',\tau_3';T)
\nonumber\\  &&\!\!\!\!\!\!\!\!
=\pathint{\tau_1}\pathint{\tau_2}\pathint{\tau_3}
\bigg(\frac{1}{\cosh^2\tau_1}-\frac{1}{\cosh^2\tau_2}\bigg)
\frac{1}{\cosh\tau_1\cosh\tau_2}
\nonumber\\  &&\!\!\!\!\!\!\!\!\qquad\times
\exp\left\{\ih\int_0^T\Bigg[\frac{m}{2}
\left(\frac{\cosh^2\tau_2-\cosh^2\tau_1}{\cosh^2\tau_1\cosh^2\tau_2}
(\dot\tau_1^2-\dot\tau_2^2)+\frac{\dot\tau_3^2}{\cosh^2\tau_1\cosh^2\tau_2}
       \right)-\frac{3\hbar^2}{8m}\Bigg]\dt\right\}\enspace.\qquad\quad
\end{eqnarray}
The path integral for this coordinate system corresponds to the path integral
for the first elliptic parabolic system on
the three-dimensional hyperboloid in its complexified form. We have from
\cite{GROad} the path integral representation 
($a>0$, $|\vtheta|<\pi/2,\vrho\in\bbbr$):
\begin{eqnarray}
& &\!\!\!\!\!\!\!\!\!\!
  \pathint{a}\pathint\vtheta
   {\cosh^2a-\cos^2\vtheta\over\cosh^3a\cos^3\vtheta}\pathint\vrho
         \nonumber\\   & &\!\!\!\!\!\!\!\!\!\!\qquad\times
   \exp\Bigg[{\i m\over2\hbar}\intt
   {(\cosh^2a-\cos^2\vtheta)(\dot a^2+\dot\vtheta^2)+\dot\vrho^2\over
    \cosh^2a\cos^2\vtheta}dt-{3\i\hbar T\over 8m}\Bigg]
         \nonumber\\   & &\!\!\!\!\!\!\!\!\!\!
   =\sqrt{\cosh a'\cosh a''}\,\cos\vtheta'\cos\vtheta''
    \int_{\bbbr}{dk_\vrho\over2\pi}\e^{\i k_\vrho(\vrho''-\vrho')}
         \nonumber\\   & &\!\!\!\!\!\!\!\!\!\!\qquad\times
    \int_0^\infty dp\,\sinh\pi p
    \int_0^\infty{dk\,k\sinh\pi k\over(\cosh^2\pi k+\sinh^2\pi p)^2}
    \,\Energyldrei
         \nonumber\\   & &\!\!\!\!\!\!\!\!\!\!\qquad\times
   \sum_{\epsilon,\epsilon'=\pm1}
   S_{\i p-1/2}^{\i k\,(1)}(\epsilon\tanh a'';\i k_\vrho)
   S_{\i p-1/2}^{\i k\,(1)\,*}(\epsilon\tanh a';\i k_\vrho)
         \nonumber\\   & &\!\!\!\!\!\!\!\!\!\!\qquad\times
   \ps_{\i k-1/2}^{\i p}(\epsilon'\sin\vtheta'';-k_\vrho^2)
   \ps_{\i k-1/2}^{\i p\,*}(\epsilon'\sin\vtheta';-k_\vrho^2)\enspace.
\label{Solution-16}
\end{eqnarray}
The $\ps_\mu^\nu$ and $S_{\mu}^{\nu\,(1)}$ are spheroidal wave-functions. 
We can achieve the connection by the coordinate substitution
$\vtheta=\i\tau_2$ and $\vrho=\tau_3$.
Equation (\ref{Solution-16}) is actually the solution to the original problem,
with continuous spectrum as in the previous case,
because its formulation comes from the complexification of the corresponding
coordinate system on the three-dimensional hyperboloid, and we are done.

\subsection{System 16: Parabolic}
\message{System 16: Parabolic}
This coordinate system is defined as
\begin{equation}
\left.\begin{array}{ll}
\displaystyle
 z_1=\frac{(\xi^2+\eta^2)^2+4}{8\xi\eta}+\frac{\tau^2}{2\xi\eta}\qquad
&\displaystyle
z_2=-\i\frac{\tau}{\xi\eta}
\\[3mm]
\displaystyle
 z_3=-\frac{\i}{2}\bigg(\frac{\xi}{\eta}-\frac{\eta}{\xi}\bigg) 
&\displaystyle
z_4=\i\bigg(\frac{(\xi^2+\eta^2)^2-4}{8\xi\eta}+\frac{\tau^2}{2\xi\eta}\bigg)
\end{array}\quad\right\}
\end{equation}
($\xi,\eta>0,\tau\in\bbbr$). The set of commuting operators is given by
\begin{equation}
\CL_1=(I_{42}+\i I_{21})^2,\qquad 
\CL_2=\{I_{32},I_{42}+\i I_{21}\}-\{I_{41},\i I_{34}-I_{31}\}\enspace.
\end{equation}
The metric terms are
\begin{equation}
\left.\begin{array}{l}
\displaystyle
\d s^2=\bigg(\frac{1}{\xi^2}+\frac{1}{\eta^2}\bigg)
            (\d\xi^2+\d\eta^2)+\frac{1}{\xi^2\eta^2}\d\tau^2\enspace,
\\[3mm]
\displaystyle
\sqrt{g}=\bigg(\frac{1}{\xi^2}+\frac{1}{\eta^2}\bigg)\frac{1}{\xi\eta}\enspace,
\\
\Gamma_\xi=-\dfrac{2/\xi^3}{1\xi^2+1/\eta^2}-\dfrac{1}{\xi},\quad
\Gamma_\eta=-\dfrac{2/\eta^3}{1/\xi^2+1/\eta^2}-\dfrac{1}{\eta},\quad
\Gamma_\tau=0\enspace.
\end{array}\quad\right\}
\end{equation}
We have for the momentum operators
\begin{eqnarray}
p_\xi&=&=\hi\bigg(\Partial{\xi}+\frac{\xi}{\xi^2+\eta^2}-\frac{3}{2\xi}\bigg)
\enspace,
\\
p_\eta&=&=\hi\bigg(\Partial{\eta}+
          \frac{\eta}{\xi^2+\eta^2}-\frac{3}{2\eta}\bigg)\enspace,
\\
p_\tau&=&\hi\Partial{\tau}\enspace,
\end{eqnarray}
and the Hamiltonian is given by
\begin{eqnarray}
H&=&-\frac{\hbar^2}{2m}\Bigg[\frac{\xi^2\eta^2}{\xi^2+\eta^2}
\bigg(\Partialsq{\xi}-\frac{1}{\xi}\Partial{\xi}
+\Partialsq{\eta}-\frac{1}{\eta}\Partial{\eta}\bigg)
+\xi^2\eta^2\Partialsq{\tau}\Bigg]
\nonumber\\
&=&\frac{1}{2m}\left[\sqrt{\frac{\xi^2\eta^2}{\xi^2+\eta^2}}
\,\big(p_\xi^2+p_\eta^2\big)\sqrt{\frac{\xi^2\eta^2}{\xi^2+\eta^2}}
+\xi^2\eta^2\,p_\tau^2\right]+\frac{3\hbar^2}{8m}\enspace.
\end{eqnarray}
For the path integral we find by separating off the $\tau$-path integration
(plane waves)
\begin{eqnarray}
&&K^{(\SdreiC)}(\xi'',\xi',\eta'',\eta',\tau'',\tau';T)
\nonumber\\  &&
=\pathint{\xi}\pathint{\eta}\pathint{\tau}\frac{\xi^2+\eta^2}{\xi^3\eta^3}
\nonumber\\  &&\qquad\times
\exp\left\{\ih\int_0^T\Bigg[\frac{m}{2}
\left(\frac{\xi^2+\eta^2}{\xi^2\eta^2}
      (\dot\xi^2+\dot\eta^2)+\frac{\dot\tau^2}{\xi^2\eta^2}\right)
-\frac{3\hbar^2}{8m}\Bigg]\dt\right\}
\nonumber\\  &&
=(\xi'\xi''\eta'\eta'')^{1/2}
\int_{\bbbr}\d k\frac{\e^{\i k(\tau''-\tau')}}{2\pi}
\pathint{\xi}\pathint{\eta}\frac{\xi^2+\eta^2}{\xi^2\eta^2}
\nonumber\\  &&\qquad\times
\exp\left\{\ih\int_0^T\Bigg[\frac{m}{2}\frac{\xi^2+\eta^2}{\xi^2\eta^2}
(\dot\xi^2+\dot\eta^2)-\frac{\hbar^2k^2}{2m}\xi^2\eta^2
-\frac{3\hbar^2}{8m}\Bigg]\dt\right\}\qquad\qquad
\nonumber\\  &&
=(\xi'\xi''\eta'\eta'')^{1/2}
\int_{\bbbr}\d k\frac{\e^{\i k(\tau''-\tau')}}{2\pi}
\int_{\bbbr}\frac{\d E}{2\pi\hbar}\e^{\i ET/\hbar}\int_0^\infty \d s''
K_k(\xi'',\xi',\eta'',\eta';s'')\enspace,
\end{eqnarray}
with the transformed path integral $K_k(s'')$ given by:
\begin{eqnarray}
&&K_k(\xi'',\xi',\eta'',\eta';s'')
=\pathints{\xi}\pathints{\eta}
\nonumber\\  &&\qquad\times
\exp\left\{\ih\int_0^{s''}\left[\frac{m}{2}\bigg((\dot\xi^2+\dot\eta^2)
-\omega^2(\xi^2+\eta^2)\bigg)
-\hbar^2\frac{\lambda^2-\viert}{2m}\bigg(\frac{1}{\xi^2}+\frac{1}{\eta^2}\bigg)
\right]\d s\right\}\enspace.\qquad
\end{eqnarray}
I have set $\omega=\hbar|k|/m$ and $\lambda=\pm\sqrt{1+2mE/\hbar^2}$.
Each of the two path integrals in $\xi$ and $\eta$ are path integrals
for the radial harmonic oscillator. We insert the expansions into the
discrete wave-functions and obtain
\begin{eqnarray}
&&K_k(\xi'',\xi',\eta'',\eta';s'')
=\frac{4m^2\omega^2}{\hbar^2}\sqrt{\xi'\xi''\eta'\eta''}
\nonumber\\  &&\qquad\times
\sum_{n_\xi,n_\eta}
\frac{n_\xi!}{\Gamma(n_\xi+\lambda+1)}
\frac{n_\eta!}{\Gamma(n_\eta+\lambda+1)}
\bigg(\frac{m^2\omega^2}{\hbar^2}\xi'\xi''\eta'\eta''\bigg)^\lambda
\e^{-2\i\omega(n_\xi+n_\eta+\lambda+1)s''}
\nonumber\\  &&\qquad\times
L_{n_\xi} ^{(\lambda)}\bigg(\frac{m\omega}{\hbar}{\xi'}^2\bigg)
L_{n_\xi} ^{(\lambda)}\bigg(\frac{m\omega}{\hbar}{\xi''}^2\bigg)
L_{n_\eta}^{(\lambda)}\bigg(\frac{m\omega}{\hbar}{\eta'}^2\bigg)
L_{n_\eta}^{(\lambda)}\bigg(\frac{m\omega}{\hbar}{\eta''}^2\bigg)
\nonumber\\  &&\qquad\times
\exp\bigg[-\frac{m\omega}{2\hbar}
          ({\xi'}^2+{\xi''}^2+{\eta'}^2+{\eta''}^2)\bigg]\enspace.
\end{eqnarray}
Since $G(E)=\ih\int_0^\infty K(s'')\d s''$ we obtain by taking the minus sign
in the square-root of $\lambda$ by performing the $s''$-integration the energy-levels
\begin{equation}
E_{n_\xi n_\eta}=\frac{\hbar^2}{2m}(n_\xi+n_\eta)(n_\xi+n_\eta+2)
\equiv \frac{\hbar^2}{2m}J(J+2)\enspace,
\end{equation}
with $J=n_\xi+n_\eta$. Evaluating the residua and ordering factors
therefore yields
\begin{equation}
K^{(\SdreiC)}(\xi'',\xi',\eta'',\eta',\tau'',\tau';T)
=\int_{\bbbr}\sum_{n_\xi,n_\eta}
\Psi_{kn_\xi n_\eta}(\xi',\eta',x')\Psi_{kn_\xi n_\eta}^*(\xi'',\eta'',x'')
\e^{-\i E_jT/\hbar}
\end{equation}
with the wave-functions  on $\SdreiC$ given by
\begin{eqnarray}
\Psi_{kn_\xi n_\eta}(\xi,\eta,\tau)&=&
\frac{\e^{\i k\tau}}{\sqrt{2\pi}}
\sqrt{\hbar|k| \frac{\xi n_\xi!}{\Gamma(n_\xi+\lambda+1)}
\frac{\eta n_\eta!}{\Gamma(n_\eta+\lambda+1)}}\,
\bigg(\frac{m^2\omega^2}{\hbar^2}\xi\eta\bigg)^{J+1}
\nonumber\\  &&\qquad\times
L_{n_\xi} ^{(J+1)}\bigg(\frac{m\omega}{\hbar}{\xi}^2\bigg)
L_{n_\eta}^{(J+1)}\bigg(\frac{m\omega}{\hbar}{\eta}^2\bigg)
\exp\bigg[-\frac{m\omega}{2\hbar}(\xi^2+\eta^2)\bigg]\enspace,\qquad
\end{eqnarray}
and the energy-spectrum (\ref{Energy-J}). 
This concludes the discussion.

\subsection{System 17: Ellipsoidal}
\message{System 17: Ellipsoidal I}
This coordinate system is defined as
\begin{equation}
\left.\begin{array}{ll}
\displaystyle
 z_1^2=-\frac{\vrho_1\vrho_2\vrho_3}{ab}
&\displaystyle
 z_2^2=\frac{(\vrho_1-1)(\vrho_2-1)(\vrho_3-1)}{(a-1)(b-1)}
\\[3mm]
\displaystyle
 z_3^2=-\frac{(\vrho_1-b)(\vrho_2-b)(\vrho_3-b}{(a-b)(b-1)b}  \qquad       
&\displaystyle
 z_4^2=\frac{(\vrho_1-a)(\vrho_2-a)(\vrho_3-a)}{(a-b)(a-1)a}\enspace.
\end{array}\quad\right\}
\end{equation}
The set of commuting operators is given by
\begin{equation}
\left.\begin{array}{l}
\CL_1=abI_{12}^2+aI_{13}^2+bI_{14}^2\enspace,\\[3mm]
\CL_2=(a+b)I_{12}^2+(a+1)I_{13}^2+(b+1)I_{14}^2
         +aI_{32}^2+bI_{42}^2+I_{43}^2\enspace.
\end{array}\quad\right\}
\end{equation}
The metric terms are given by
\begin{equation}
\d s^2=\frac{(\vrho_1-\vrho_2)(\vrho_1-\vrho_3)}{f(\vrho_1)}\d\vrho_1^2
      +\frac{(\vrho_2-\vrho_3)(\vrho_2-\vrho_1)}{f(\vrho_2)}\d\vrho_2^2
      +\frac{(\vrho_3-\vrho_1)(\vrho_3-\vrho_2)}{f(\vrho_3)}\d\vrho_3^2
\enspace,
\end{equation}
with $f(\vrho)=-4(\vrho-a)(\vrho-b)(\vrho-1)\vrho$. 
For the path integral we find
\begin{eqnarray}
&&K^{(\SdreiC)}(\vrho_1'',\vrho_1',\vrho_2'',\vrho_2',\vrho_3'',\vrho_3';T)
=\pathint{\vrho_1}\pathint{\vrho_2}\pathint{\vrho_3}
\nonumber\\  &&\qquad\times
\exp\Bigg\{\ih\int_0^T\Bigg[\frac{m}{2}
\frac{(\vrho_1-\vrho_2)(\vrho_1-\vrho_3)}{f(\vrho_1)}\dot \vrho_1^2
      +\frac{(\vrho_2-\vrho_3)(\vrho_2-\vrho_1)}{f(\vrho_2)}\dot \vrho_2^2
\nonumber\\  &&\qquad\qquad\qquad\qquad\qquad\qquad
      +\frac{(\vrho_3-\vrho_1)(\vrho_3-\vrho_2)}{f(\vrho_3)}\dot \vrho_3^2
-\Delta V(\vrho_1,\vrho_2,\vrho_3)\Bigg]\dt\Bigg\}\enspace.\qquad\qquad
\end{eqnarray}
It is obvious that such a path integral in ellipsoidal coordinates is
not tractable. We let the result as it stands, and the same statement
is valid for the remaining ``ellipsoidal''-related coordinate systems.
Let us, however, note that we can state the propagator in a formal way
by a construction it 
from the wave-functions according to \cite{GROad}. We have found
the following representation on the real sphere $S^{(3)}$:
\begin{eqnarray}
&&  \pathint{\vrho_1}\pathint{\vrho_2}\pathint{\vrho_3}
   {(\vrho_2-\vrho_1)(\vrho_3-\vrho_2)(\vrho_3-\vrho_1)\over
             8\sqrt{P(\vrho_1)P(\vrho_2)P(\vrho_3)}}\qquad\qquad\qquad\qquad
         \nonumber\\   & &\qquad\qquad\times
  \exp\Bigg\{\ih\intt\Bigg[{m\over2}
    \sum_{i=1}^3g_{\vrho_i\vrho_i}\dot\vrho_i^2
   -\Delta V_{PF}(\vec\vrho\,)\Bigg]dt\Bigg\}
         \nonumber\\   & &
  =\sum_{2s=0}^\infty\sum_{\lambda,\mu}\e^{-2\i\hbar Ts(s+1)/m}
  \Psi_{s,\lambda,\mu}^*(\vrho'_1,\vrho'_2,\vrho'_3)
  \Psi_{s,\lambda,\mu}(\vrho''_1,\vrho''_2,\vrho''_3)
\end{eqnarray}
with ellipsoidal coordinates defined on  $S^{(3)}$
($d<\vrho_3<c<\vrho_2<b<\vrho_1<a$):
\begin{equation}
\left.\begin{array}{l}
\displaystyle
s_1^2=\dfrac{(\vrho_1-d)(\vrho_2-d)(\vrho_3-d)}{(a-d)(b-d)(c-d)}\\[3mm]
\displaystyle
s_2^2=\dfrac{(\vrho_1-c)(\vrho_2-c)(\vrho_3-c)}{(a-c)(b-c)(d-c)}\\[3mm]
\displaystyle
s_3^2=\dfrac{(\vrho_1-a)(\vrho_2-a)(\vrho_3-a)}{(d-a)(c-a)(b-a)}\\[3mm]
\displaystyle
s_4^2=\dfrac{(\vrho_1-b)(\vrho_2-b)(\vrho_3-b)}{(d-b)(c-b)(a-b)}\enspace.
\end{array}\ \right\}
\end{equation}
The metric tensor in ellipsoidal coordinates then has the form
\begin{equation}
  (g_{ab})=-\viert\diag\bigg(
  {(\vrho_1-\vrho_2)(\vrho_1-\vrho_3)\over P(\vrho_1)},
  {(\vrho_2-\vrho_3)(\vrho_2-\vrho_1)\over P(\vrho_2)},
  {(\vrho_3-\vrho_1)(\vrho_3-\vrho_2)\over P(\vrho_3)}\bigg)\enspace,
\end{equation}
and $P(\vrho)=(\vrho-a)(\vrho-b)(\vrho-c)(\vrho-d)$.
Here, we have adopted the notation of Karayan et al.\ \cite{AKPSZ,GKPSa}.  
The quantum numbers are the eigenvalues of the operators which characterize the
ellipsoidal system on the sphere, thus giving a complete set of
observables corresponding to the ellipsoidal coordinates on the
sphere. However, this representation remains on a formal level
and the corresponding wave-functions are explicitly know only on the
real three-dimensional sphere. 
For details see \cite{AKPSZ,GKPSa,LUKAb,KMW,LUNA,KOKU}.
The ellipsoidal system exists on the three-dimensional sphere (System VI.) 
on the three-dimensional hyperboloid (System XVXIII.). 

\subsection{System 18}
\message{System 18}
For the remaining coordinate system we state only their definition.
The corresponding quantum theory setup is formally the same as for the 
Ellipsoidal coordinates, however with a different function $f(\vrho)$.

This coordinate system is defined as
\begin{equation}
\left.\begin{array}{ll}
\displaystyle
 (\i z_1+z_2)^2=\frac{\vrho_1\vrho_2\vrho_3}{a}
&\displaystyle
 z_1^2+z_2^2=\frac{1}{a^2}
\Big[(a+1)\vrho_1\vrho_2\vrho_3-a(\vrho_1\vrho_2+\vrho_1\vrho_3
+\vrho_2\vrho_3)\big]
\\[3mm]
\displaystyle
 z_3^2=\frac{(\vrho_1-1)(\vrho_2-1)(\vrho_3-1)}{1-a}
&\displaystyle
z_4^2=\frac{(\vrho_1-a)(\vrho_2-a)(\vrho_3-a)}{a^2(a-1)}
\end{array}\ \right\}
\end{equation}
The set of commuting operators is given by
\begin{equation}
\left.\begin{array}{l}
\CL_1=(I_{42}-\i I_{14})^2-a(I_{32}+\i I_{13})^2-aI_{12}^2\enspace,\\[2mm]
\CL_2=(a+1)I_{12}^2+I_{14}^2+I_{42}^2-a(I_{13}^2+I_{32}^2)
+(I_{42}+\i I_{14})^2+(I_{32}+\i I_{13})^2\enspace.
\end{array}\quad\right\}
\end{equation}
The metric terms are given by
\begin{equation}
\d s^2=\frac{(\vrho_1-\vrho_2)(\vrho_1-\vrho_3)}{f(\vrho_1)}\d \vrho_1^2
      +\frac{(\vrho_2-\vrho_3)(\vrho_2-\vrho_1)}{f(\vrho_2)}\d \vrho_2^2
      +\frac{(\vrho_3-\vrho_1)(\vrho_3-\vrho_2)}{f(\vrho_3)}\d \vrho_3^2
\enspace,
\end{equation}
with $f(\vrho)=-4(\vrho-2)(\vrho-1)\vrho^2$.
Systems 18 corresponds to Systems 31--33 on the three-dimensional hyperboloid.

\subsection{System 19}
\message{System 19}
This coordinate system is defined as
\begin{equation}
\left.\begin{array}{ll}
 (z_1+\i z_2)^2=-(\vrho_1-1)(\vrho_2-1)(\vrho_3-1)
&z_1^2+z_2^2=2\vrho_1\vrho_2\vrho_3
-(\vrho_1\vrho_3+\vrho_2\vrho_3+\vrho_1\vrho_2)+1
\\[2mm]
 (z_3+\i z_4)^2=-\vrho_1\vrho_2\vrho_3
&z_3^2+z_4^2=\vrho_1\vrho_3+\vrho_2\vrho_3+\vrho_1\vrho_2
-2\vrho_1\vrho_2\vrho_3\enspace.
\end{array}\ \right\}
\end{equation}
The set of commuting operators is given by
\begin{equation}
\left.\begin{array}{l}
\CL_1=2(I_{31}+\i I_{32})^2+\{I_{31}+\i I_{32},I_{24}+\i I_{41}\}+I_{12}^2\enspace,
\\[2mm]
\CL_2=2(I_{31}+\i I_{32})^2+\{I_{31}+\i I_{32},I_{24}+I_{41}\}-I_{34}^2\enspace.
\end{array}\ \right\}
\end{equation}
The metric terms are given by
\begin{equation}
\d s^2=\frac{(\vrho_1-\vrho_2)(\vrho_1-\vrho_3)}{f(\vrho_1)}\d \vrho_1^2
      +\frac{(\vrho_2-\vrho_3)(\vrho_2-\vrho_1)}{f(\vrho_2)}\d \vrho_2^2
      +\frac{(\vrho_3-\vrho_1)(\vrho_3-\vrho_2)}{f(\vrho_3)}\d \vrho_3^2
\enspace,
\end{equation}
with $f(\vrho)=-4(\vrho-1)^2\vrho^2$.

\subsection{System 20}
\message{System 20}
This coordinate system is defined as
\begin{equation}
\left.\begin{array}{l}
(z_2-\i z_1)^2z_1^2+z_2^2+z_3^2=\vrho_1\vrho_2\vrho_3\\[2mm]
-2z_3(z_2-\i z_1)
=\vrho_1\vrho_2+\vrho_1\vrho_3+\vrho_2\vrho_3-\vrho_1\vrho_2\vrho_3
\\[2mm]
z_1^2+z_2^2+z_3^2
=\vrho_1\vrho_2\vrho_3-\vrho_1\vrho_2-\vrho_1\vrho_3
-\vrho_2\vrho_3+\vrho_1+\vrho_2+\vrho_3\\[2mm]
z_4^2=-(\vrho_1-1)(\vrho_2-1)(\vrho_3-1)\enspace.
\end{array}\quad\right\}
\end{equation}
The set of commuting operators is given by
\begin{equation}
\left.\begin{array}{l}
\CL_1=(I_{41}+\i I_{42})^2+\{I_{32}-\i I_{13},I_{12}\}\enspace,\\[2mm]
\CL_2=I_{41}^2+I_{42}^2-I_{34}^2
     -(I_{41}+\i I_{42})^2+\{I_{41}-\i I_{42},I_{34}\}\enspace.
\end{array}\quad\right\}
\end{equation}
The metric terms are given by
\begin{equation}
\d s^2=\frac{(\vrho_1-\vrho_2)(\vrho_1-\vrho_3)}{f(\vrho_1)}\d \vrho_1^2
      +\frac{(\vrho_2-\vrho_3)(\vrho_2-\vrho_1)}{f(\vrho_2)}\d \vrho_2^2
      +\frac{(\vrho_3-\vrho_1)(\vrho_3-\vrho_2)}{f(\vrho_3)}\d \vrho_3^2
\enspace,
\end{equation}
with $f(\vrho)=-4(\vrho-1)\vrho^3$.
Systems 20 corresponds to System 34 on the three-dimensional hyperboloid.

\subsection{System 21}
\message{System 21}
This coordinate system is defined as
\begin{equation}
\left.\begin{array}{l}
(z_1+\i z_2)^2=2\vrho_1\vrho_2\vrho_3\\[2mm]
(z_1+\i z_2)((z_3+\i z_4)=-(\vrho_1\vrho_2+\vrho_2\vrho_3+\vrho_1\vrho_3)
\\[2mm]
 -(z_1+\i z_2)(z_3-\i z_4)+\bhalf(z_3+\i z_4)^2=\vrho_1+\vrho_2+\vrho_3\enspace.
\end{array}\ \right\}
\end{equation}
The set of commuting operators is given by
\begin{equation}
\left.\begin{array}{l}
\CL_1=\bhalf\{I_{21},I_{41}+I_{23}+\i(I_{31}+I_{24})\}
     -\bviert\big[I_{13}+I_{24}+\i(I_{23}+I_{41})\big]^2\enspace,\\[2mm]
\CL_2=\bhalf\{I_{21}+I_{43},I_{32}+I_{14}+\i(I_{13}+I_{24})\}\\[2mm]
\qquad+\bhalf\{I_{41}+I_{23}+\i(I_{31}+I_{24}),I_{43}\}
+\bhalf (I_{42}+\i I_{23})^2-\bhalf(I_{13}+\i I_{14})^2\enspace.
\end{array}\quad\right\}
\end{equation}
The metric terms are given by
\begin{equation}
\d s^2=\frac{(\vrho_1-\vrho_2)(\vrho_1-\vrho_3)}{f(\vrho_1)}\d \vrho_1^2
      +\frac{(\vrho_2-\vrho_3)(\vrho_2-\vrho_1)}{f(\vrho_2)}\d \vrho_2^2
      +\frac{(\vrho_3-\vrho_1)(\vrho_3-\vrho_2)}{f(\vrho_3)}\d \vrho_3^2
\enspace,
\end{equation}
with $f(\vrho)=-4\vrho^4$.


\section{Summary and Discussion}
\message{Summary and Discussion}
The archived results are very satisfactory. We could find many path integral
representations on the complex sphere, several of them included the
extension and application of already know results, several others of them are
completely new.
The most important result consists of the incorporation of the complex Liouville
potential into the path integral formalism.
However, we must always keep in mind that the complex sphere is an abstract space,
which means that the various path integral representations require an interpretation
depending whether one considers a compact or non-compact variable range.
In the compact case, the abstract complex space allows the interpretation of the real
three-dimensional sphere with its discrete spectrum. Here, we can identify the
six coordinate systems as indicated in Table  \ref{SdreiCcosys}: They are systems
(1), (3), (6), (13), and (17), where No.(13) is counted twice to include the
prolate as well as the oblate spheroidal cases.
In the non-compact case allows the interpretation of the three-dimensional
$\Lambda^{(3)}$ and O(2,2)-hyperboloid, respectively, with a continuous spectrum.
Therefore the eigenvalues of the complex sphere
\begin{equation}
\hbarm\sigma(\sigma+2)\longleftarrow
\left\{\begin{array}{lll}
\displaystyle\hbarm l(l+2)\qquad  &l=0,1,2,\dots  &\qquad\hbox{sphere}\\[3mm]
\displaystyle\hbarm (p^2+1)\qquad &p>0            &\qquad\hbox{hyperboloid.}
\end{array}\right.
\label{spectrum-S3C}
\end{equation}
This includes the replacement of the summation of the discrete principal quantum
number $l$, say, by the principal continuous quantum number $p$, i.e.,
$\sum_l\to\int_0^\infty \d p$.
Furthermore, the wave-functions have to analytically continued, say the discrete
wave-functions for the spherical coordinate system (3) on the real sphere:
\begin{equation}
\Psi_{J,m_1,m_2}(\chi,\vtheta,\vphi)=N^{-1/2}\e^{\i m_1\vphi}
   (\sin\chi)^{m_1}C_{J-m_1}^{m_1+2}(\cos\chi)
   (\sin\vtheta)^{m_2}C_{m_1-m_2}^{m_2+3/2}(\cos\vtheta)\enspace,
\end{equation}
must be replaced by the continuous wave-functions for the spherical
system on the hyperboloid $\Lambda^{(3)}$ ($Y_l^{m}(\vtheta,\vphi)$ are the usual
spherical harmonics on the two-dimensional sphere, c.f. Section 2.3) \cite{GROad}:
\begin{equation}
\Psi_{p,l,m}(\tau,\vtheta,\vphi)
=Y_l^{m}(\vtheta,\vphi)(\sinh\tau)^{-1/2}
  \sqrt{p\sinh\pi p\over\pi}\Gamma(\i p+l+1)
  \CP_{\i p-1/2}^{-\half-l}(\cosh\tau)\enspace.
\end{equation}
Also, the invariant distance (under rotations) on the real sphere must be replaced 
by the invariant distance on the hyperboloid, 
c.f. Eqs. (\ref{distance-sphere},\ref{distance-hyperboloid}) with the corresponding
Green functions  (\ref{GS3},\ref{Green-hyperboloid}), respectively.

In order to find the corresponding solutions on the O(2,2) hyperboloid
matters are more difficult, because in addition to a continuous spectrum also
a discrete spectrum is present. These issues will be discussed in future
publication.

In Section II, I have displayed the path integral solutions (1) to (12). They are
characterized by the property that they have a subgroup, respectively a 
subspace structure. This has also been emphasized in Table  \ref{SdreiCcosys},
where the explicit subspace with its corresponding coordinate representation
has been displayed. Remarkably, for all coordinate systems a path integral
representation could be found. The principal new representation was the one for
the horicyclic system, respectively the complex Liouville potential, i.e.
\begin{eqnarray}
&&\pathint{x}\exp\left[\ih\int_0^T
\left(\frac{m}{2}\dot x^2-\frac{\hbar^2}{2m}k^2\e^{-2\i x}
\right)\dt\right]
\nonumber\\  &&\qquad
=\sum_{J\in\bbbn_0} \bhalf
H^{(1)}_{J+1/2}\Big(k\,\e^{-\i x''}\Big)
H^{(1)}_{J+1/2}\Big(k\,\e^{-\i x'}\Big)
\exp\Bigg[-\ih\frac{\hbar^2J(J+2)}{2m}T\Bigg]\enspace.\qquad
\end{eqnarray}
This path integral representation was very useful in evaluating
several path integral representations involving other horicyclic coordinate
possibilities.

In several other coordinate systems we could exploit already known results,
for instance
from the two-dimensional sphere (e.g. sphero-elliptic (6)), 
from the two-dimensional hyperboloid (e.g. spherical-degenerate elliptic I (7)),
from the two-dimensional pseudo-Euclidean plane (e.g. horicyclic-elliptic (9)), the
Euclidean plane (e.g. horicyclic-parabolic I (11)), and others as noted in the text.

In Section III, I have discussed the path integral representations of the
non-subspace cases which are much more involved. Starting with the elliptic-cylindrical
system we can only state the formal solution as known from the three-dimensional
sphere, which can be represented by Lam\'e polynomials. Four more representations
could be explicitly stated, the last for the ellipsoidal one also only as a formal
solutions in terms of {\em ellipsoidal wave-functions} as known from the
three-dimensional sphere. For the remaining ``ellipsoidal'' systems nothing is
known which is not surprising due to their very complicated structure.

There are several three-dimensional other spaces where a path integral treatment
for various coordinate systems is possible: these are the single-sheeted hyperboloid,
the O(2,2)-hyperboloid \cite{GROad}, Darboux spaces in three dimensions \cite{GROat}, 
and Koenigs spaces in two and three dimensions \cite{GROau}. 
The latter two open the
possibility to discuss quantum motion on spaces of non-constant curvature,
whereas the former two have the property that in addition to the continuous spectrum
also an infinite discrete spectrum exists. 
In particular, the quantum motion on the O(2,2) hyperboloid is of special
interest due to the fact that on the O(2,2) hyperboloid a discrete {\it and} a 
continuous spectrum exists. Therefore the complete spectrum consists of both
contributions of (\ref{spectrum-S3C}), and almost all path integral representations
with their wave-function expansions must be analytically continued to
a discrete {\it and} a continuous contribution.
These issues will be discussed in a future publication.

\subsection*{Acknowledgments}
This work was supported by the Heisenberg--Landau program. 

I would like to thank George Pogosyan (Yerevan State University) 
for helpful discussions on the properties of coordinate systems and superintegrability. 
I also would like to thank L.Mardoyan for the warm hospitality during my stay in
Dubna, Russia, and G.Pogosyan, for the warm hospitality during my stay in
Yerevan, Armenia.


\input cyracc.def
\font\tencyr=wncyr10
\font\tenitcyr=wncyi10
\font\tencpcyr=wncysc10
\def\cyrrm{\tencyr\cyracc}
\def\cyrit{\tenitcyr\cyracc}
\def\cyrcp{\tencpcyr\cyracc}
\renewcommand{\baselinestretch}{1.05}
\footnotesize
\addcontentsline{toc}{section}{References}%

\bigskip\bigskip 
\vbox{\centerline{\ }
\centerline{\quad\epsfig{file=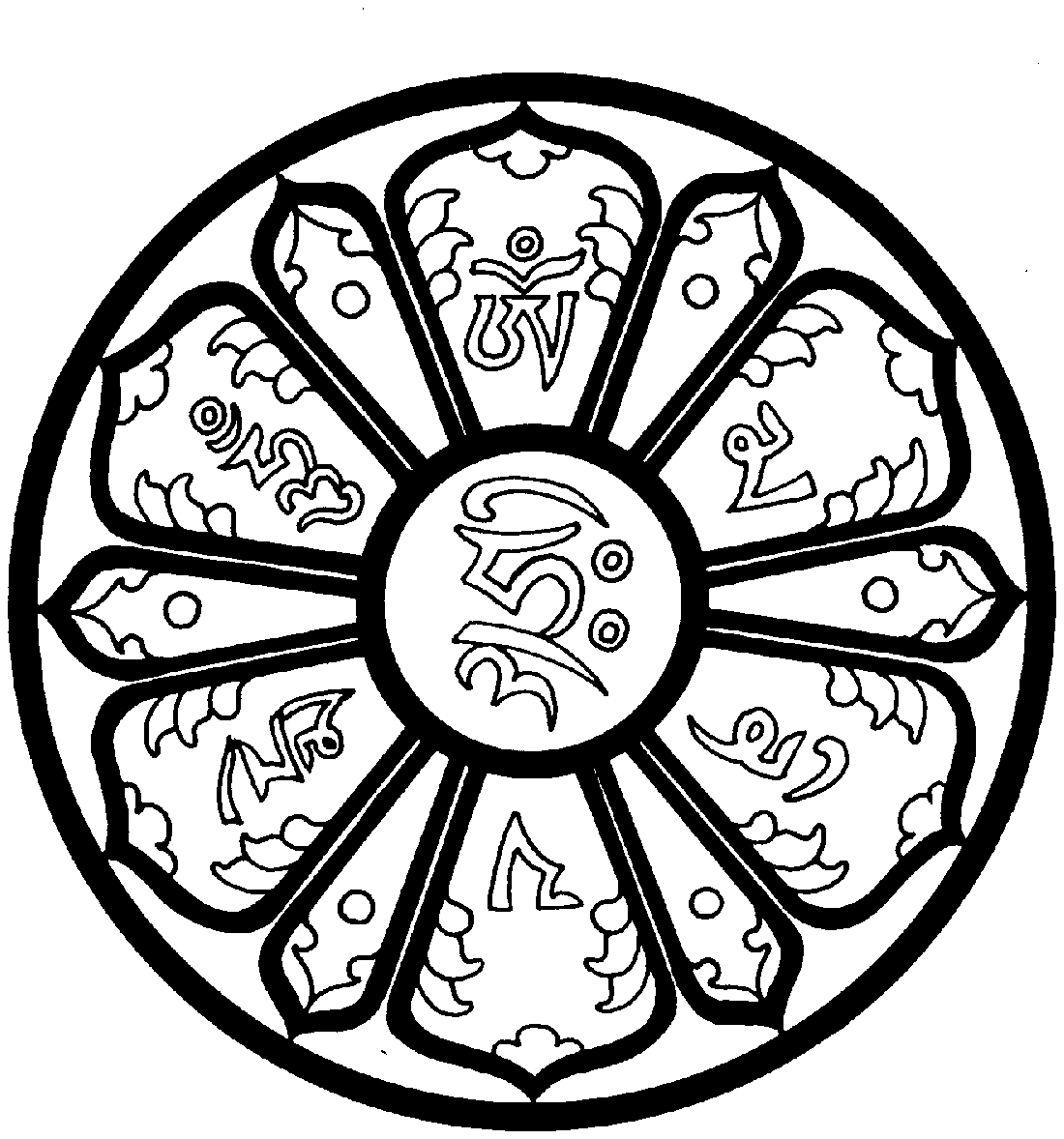,width=4cm}}}

\end{document}